\documentstyle[12pt,epsfig]{article}
\newcommand{\n}{\hspace*{-2.5mm}}
\newcommand{\li}{\mathop{{\rm Li}_2}\nolimits}
\newcommand{\simgt}{\,\rlap{\lower 3.5 pt \hbox{$\mathchar \sim$}} \raise 1pt
 \hbox {$>$}\,}
\newcommand{\simlt}{\,\rlap{\lower 3.5 pt \hbox{$\mathchar \sim$}} \raise 1pt
 \hbox {$<$}\,}

\catcode`@=11
\newcount\@tempcntc
\def\@citex[#1]#2{\if@filesw\immediate\write\@auxout{\string\citation{#2}}\fi
  \@tempcnta\z@\@tempcntb\m@ne\def\@citea{}\@cite{\@for\@citeb:=#2\do
    {\@ifundefined
       {b@\@citeb}{\@citeo\@tempcntb\m@ne\@citea\def\@citea{,}{\bf ?}\@warning
       {Citation `\@citeb' on page \thepage \space undefined}}%
    {\setbox\z@\hbox{\global\@tempcntc0\csname b@\@citeb\endcsname\relax}%
     \ifnum\@tempcntc=\z@ \@citeo\@tempcntb\m@ne
       \@citea\def\@citea{,}\hbox{\csname b@\@citeb\endcsname}%
     \else
      \advance\@tempcntb\@ne
      \ifnum\@tempcntb=\@tempcntc
      \else\advance\@tempcntb\m@ne\@citeo
      \@tempcnta\@tempcntc\@tempcntb\@tempcntc\fi\fi}}\@citeo}{#1}}
\def\@citeo{\ifnum\@tempcnta>\@tempcntb\else\@citea\def\@citea{,}%
  \ifnum\@tempcnta=\@tempcntb\the\@tempcnta\else
   {\advance\@tempcnta\@ne\ifnum\@tempcnta=\@tempcntb \else \def\@citea{--}\fi
    \advance\@tempcnta\m@ne\the\@tempcnta\@citea\the\@tempcntb}\fi\fi}
\catcode`@=12
\begin{document}
\title{\vskip-3cm{\baselineskip14pt
\centerline{\normalsize DESY 97--012\hfill ISSN~0418--9833}
\centerline{\normalsize MPI/PhT/97--009\hfill}
\centerline{\normalsize hep--ph/9702406\hfill}
\centerline{\normalsize February 1997\hfill}}
\vskip1.5cm
Coherent Description of $D^{*\pm}$ Production in $e^+e^-$ and Low-$Q^2$ $ep$ 
Collisions}
\author{J. Binnewies$^1$, B.A. Kniehl$^2$, G. Kramer$^1$\\
$^1$ II. Institut f\"ur Theoretische Physik\thanks{Supported
by Bundesministerium f\"ur Forschung und Technologie, Bonn, Germany,
under Contract 05~6~HH~93P~(5),
and by EEC Program {\it Human Capital and Mobility} through Network
{\it Physics at High Energy Colliders} under Contract
CHRX--CT93--0357 (DG12 COMA).},
Universit\"at Hamburg,\\
Luruper Chaussee 149, 22761 Hamburg, Germany\\
$^2$ Max-Planck-Institut f\"ur Physik (Werner-Heisenberg-Institut),\\
F\"ohringer Ring 6, 80805 Munich, Germany}
\date{}
\maketitle
\begin{abstract}
We present new sets of fragmentation functions for $D^{*\pm}$ mesons, both at 
leading and next-to-leading order.
They are determined by fitting LEP1 data on inclusive $D^{*\pm}$ production in
$e^+e^-$ annihilation.
In one of the sets, we take the charm-quark fragmentation function to be of
the form proposed by Peterson et al.\ and thus obtain updated values of
the $\epsilon_c$ parameter and the $c\to D^{*+}$ branching ratio.
The new fragmentation functions lead to an excellent description of other 
$e^+e^-$ data with centre-of-mass energies between 10 and 35~GeV.
They also nicely agree with recent HERA data on inclusive $D^{*\pm}$
photoproduction in $ep$ collisions, which may be considered as a test of the
universality of the fragmentation into $D^{*\pm}$ mesons.

\end{abstract}
\newpage

\section{Introduction}

Recently, the H1 \cite{1} and ZEUS \cite{2} collaborations at HERA presented
data on the differential cross section $d^2\sigma/dy_{\rm lab}\,dp_T$ of
inclusive $D^{*\pm}$ production in low-$Q^2$ collisions, where $y_{\rm lab}$
and $p_T$ are the rapidity and transverse momentum of the produced $D^{*\pm}$
mesons in the laboratory frame, respectively. 
These measurements extended up to $p_T=12$~GeV. 
These data were compared with next-to-leading order (NLO) QCD predictions in
the so-called massless-charm scheme \cite{3} and good agreement was found.

In the massless-charm scheme, the charm-quark mass $m_c$ is neglected, which
should be a reasonable approximation for $p_T\gg m_c$. 
In this approach, charm is considered to be one of the active flavours in the
same way as the lighter $u$, $d$, and $s$ quarks. 
Then, the collinear singularities corresponding to the
$\alpha_s\ln(p_T^2/m_c^2)$ terms in a scheme where the charm-quark mass is
finite and only three active flavours are taken into account, are absorbed 
into the charm-quark parton density functions (PDF's) and into the
fragmentation functions (FF's) of charm quarks into $D^{*\pm}$ mesons. 
Thus, in a NLO analysis, the following ingredients are needed: 
$(i)$ the hard-scattering cross sections for the direct- and resolved-photon
processes calculated in the massless approximation ($m_c=0$) with $n_f=4$
active flavours and with the initial- and final-state collinear singularities,
including those of the charm quark, subtracted;
$(ii)$ the PDF's of the proton and the resolved photon, where charm is treated
as a light flavour;
$(iii)$ the FF's characterizing the hadronization of the massless partons,
including the charm quark, into $D^{*\pm}$ mesons.
(Similarly, the bottom quark is treated as a massless parton above a certain
threshold, leading to $n_f=5$.)

This massless approach was originally considered in \cite{4} as a
possibility to make predictions for large-$p_T$ heavy-quark production,
and it was first applied to the production of large-$p_T$  hadrons containing
bottom quarks in $p\overline{p}$ collisions \cite{5}. 
Subsequently, it was employed to study charm-quark production in $ep$ 
\cite{7}, $\gamma p$ \cite{8}, and $\gamma\gamma$ collisions \cite{9}.

In order to obtain definite predictions for $D^{*\pm}$ production cross
sections in the massless-charm scheme, one obviously needs the PDF's of the
charm quarks in the proton and the photon, which do not enter the massive 
calculation, as well as the FF's for $D^{*\pm}$ mesons.
Both the PDF's and FF's are basically non-perturbative input and must be
determined by experiment.

This offers the possibility to use the data of $D^{*\pm}$-photoproduction
experiments in regions where the massless-charm approach is a good
approximation, to obtain information on the charm PDF's of the proton and the
photon and on the FF's of the quarks, in particular the charm quark,
and the gluon into $D^{*\pm}$ mesons.

The FF's of the $D^{*\pm}$ mesons can be determined from measured cross
sections of $D^{*\pm}$ production in $e^+e^-$  annihilation. 
In these experiments, the centre-of-mass (CM) energy $\sqrt s$ of the
electron-positron system is large compared to $m_c$, so that the 
massless-charm approach is well justified in this case.
Then, one is left with the task of constraining the charm components of the
proton and the photon.
The photon PDF's enter, for instance, the resolved-photon cross section of
inclusive $D^{*\pm}$ photoproduction in $ep$ collisions. 
In a previous work by two of us with Spira \cite{3}, it was found that, at
moderate $p_T$, the resolved-photon cross section is of the same order of
magnitude as the direct-photon cross section.
Furthermore, the additional charm component in the proton turned out to only
contribute marginally to the cross section of inclusive $D^{*\pm}$
photoproduction.
Therefore, the study of this cross section offers an ideal possibility to
specifically learn about the charm component of the photon, on which not very
much is known from other experiments.
The traditional place to investigate the charm PDF of the photon has been
fully inclusive deep inelastic charmed-hadron production in $e\gamma$
collisions. 
Unfortunately, very little is known experimentally so far \cite{10}.

Several NLO parameterizations of the charm component of the photon exist in
the literature \cite{11,12,13}, and we shall use them for our predictions in
the following. 
The point to be made is that these charm PDF's have not yet really been tested
against experimental data which are particularly sensitive to it.

In the previous work on $D^{*\pm}$ production \cite{3}, we considered several
approximations for the fragmentation of charm quarks into $D^{*\pm}$ mesons,
including also the $Q^2$ evolution to higher scales.
The most realistic description was by the Peterson FF \cite{14} with 
evolution, which was also used for the comparison with the H1 \cite{1} and
ZEUS \cite{2} data. 
Apart from the overall normalization, the distribution of Peterson et al.,
which is used as input at the starting scale $\mu_0$, depends only on one
 parameter, $\epsilon_c$. 
The value of $\epsilon_c$ was taken from the work by Chrin \cite{15}, which is
a phenomenological analysis of data on charmed-meson production (not
necessarily restricted to $D^{*\pm}$ production) in the PETRA-PEP energy range
of $e^+e^-$  annihilation.
In the meantime, much better data, specifically on $D^{*\pm}$ production,
have been made available by the ALEPH \cite{16} and OPAL \cite{17}
collaborations at LEP1.
These data should give us an excellent handle on the $D^{*\pm}$ FF's, so that
we need not rely on Chrin's value for the $\epsilon_c$ parameter any more.

It is the purpose of this work to improve the description of the charm and 
bottom FF's into $D^{*\pm}$ mesons by fitting to the ALEPH and OPAL data on
the basis of NLO evolution equations.
These FF's will be tested against $D^{*\pm}$ production data at lower CM 
energies \cite{18} coming from PETRA \cite{30}, PEP \cite{27}, and DORIS
\cite{19}. 
They will also be used to make theoretical predictions for inclusive
$D^{*\pm}$ photoproduction at HERA, following the earlier work \cite{3}.

The outline of this paper is as follows. 
In Section~2, we shall briefly describe the NLO formalism of charm-quark
fragmentation including the transition from massless to massive
factorization of final-state collinear singularities also
employed in \cite{3}.
In this section, we shall specify the $D^{*\pm}$ FF's at the starting scale
$\mu_0$, adopting two different functional forms.
We shall consider not only charm-quark fragmentation, but also bottom-quark
fragmentation which is needed to describe the LEP1 data.
Finally, we shall present the resulting NLO predictions for $D^{*\pm}$
production in $e^+e^-$ annihilation and compare them with data from ALEPH and
OPAL used in the fits, and with the other $e^+e^-$ data at lower CM energies. 
In Section~3, we shall confront our predictions for $ep\to D^{*\pm}+X$ based 
on the two forms of FF's with the ZEUS data \cite{2}. 
Our conclusions are summarized in Section~4.

\section{\boldmath{$D^{*\pm}$} Production in \boldmath{$e^+e^-$} Annihilation}

To construct LO and NLO sets of FF's for $D^{*\pm}$ mesons, we make use of the
new LEP1 data from ALEPH \cite{16} and OPAL \cite{17}. 
In hadronic $Z$-boson decays, charmed mesons are expected to be produced
either directly through the hadronization of charm quarks in the process
$Z\rightarrow c\bar c$ or via the weak decays of $B$ hadrons produced in 
$Z\rightarrow b\bar b$, with an approximately equal rate. 
Charmed mesons from $Z\rightarrow c\bar c$ allow us to measure the charm-quark
FF.
The main task of both experimental analyses was to disentangle these two main
sources of $D^{*\pm}$ production at the $Z$-boson resonance. 
In the two papers \cite{16,17}, a number of different techniques are used to
tag $D^{*\pm}$ mesons from $B$ decay. 
The $D^{*\pm}$ mesons are reconstructed in the decay chain
$D^{*\pm}\rightarrow D^0\pi^+$ and $D^0 \rightarrow K^-\pi^+$,
which allows for a particularly clean signal reconstruction.

The results of the two measurements are separate, differential yields for
$D^{*\pm}$ mesons from $Z\to c \bar c$ and $Z\to b\bar b$ decays, normalized
to the total number of multihadronic $Z$-boson decays, as a function of
$x=2E(D^{*\pm})/\sqrt s$, where $E(D^{*\pm})$ is the measured energy of the
$D^{*\pm}$ candidate. 
The contribution due to charm-quark fragmentation is peaked at large $x$,
whereas the one due to bottom-quark fragmentation has its maximum at small
$x$. 
For the fitting procedure, we use the $x$ bins contained within the interval
$[0.1,0.9]$ and integrate the theoretical functions over the bin widths, which
is equivalent to the experimental binning procedure.
As in the experimental analyses, we sum over $D^{*+}$ and $D^{*-}$ mesons.
As a consequence, there is no difference between the FF's of a given quark
and its antiquark.

We take the starting scales for the $D^{*\pm}$ FF's of the gluon and the $u$,
$d$, $s$, and $c$ quarks and antiquarks to be $\mu_0=2m_c$, while we take
$\mu_0=2m_b$ in the case of bottom.
The FF's of the gluon and the first three flavours are assumed to be zero at
the starting scale.
These FF's are generated through the $\mu^2$ evolution, and the FF's of the
first three quarks and antiquarks coincide with each other at all scales $\mu$.
For the parameterization of the charm- and bottom-quark FF's at their
respective starting scales, we employ two different forms:

$(i)$ In our standard set (S), the FF's of the charm and bottom quarks are
both assumed to be of a form usually used for the FF's of light hadrons,
namely,
\begin{equation}
\label{standard}
D_Q(x,\mu_0^2)=Nx^{\alpha}(1-x)^{\beta},
\end{equation}
where $Q=c,b$.
Here and in the following, we skip the index $D^{*\pm}$, since we only
consider $D^{*\pm}$ production throughout this paper, e.g., $D_c$ denotes the
FF of the charm quark into $D^{*\pm}$ mesons, etc.

$(ii)$ In the so-called mixed set (M), we use for the bottom-quark FF the form
of (\ref{standard}) and for the charm-quark FF the Peterson distribution
\cite{14},
\begin{equation}
\label{peterson}
D_c(x,\mu_0^2)=N\frac{x(1-x)^2}{[(1-x)^2+\epsilon_c x]^2}.
\end{equation}
The Peterson form is particularly suitable to describe FF's that peak at large
$x$. 
Since the bottom-quark FF has its maximum at smaller $x$ values, we obtain
intolerably bad fits if we also use (\ref{peterson}) in the case of bottom.
This is the reason for choosing the bottom-quark FF at $\mu_0=2m_b$ in set M
to be of form~(\ref{standard}).

Obviously, the functional form~(\ref{standard}) has more flexibility, since,
apart from the normalization factor $N$, it depends on two parameters,
$\alpha$ and $\beta$, whereas there is just a single parameter, $\epsilon_c$,
in the Peterson distribution.

The cross section of inclusive $D^{*\pm}$ production in $e^+e^-$ annihilation,
\begin{equation}
\label{eeproc}
e^+e^-\rightarrow(\gamma,Z)\rightarrow D^{*\pm}+X,
\end{equation}
is calculated from the same formula as for any other hadron,
\begin{equation}
\label{sigx}
\frac{1}{\sigma_{\rm tot}}\,\frac{d\sigma(e^+e^-\to D^{*\pm}+X)}{dx}
=\frac{1}{\sigma_{\rm tot}}\,\sum_a\int_x^1\frac{dz}{z}
D_a\left(\frac{x}{z},M_f^2\right)
\frac{d\sigma_a}{dz}\left(z,\mu^2,M_f^2\right).
\end{equation}
Here, the sum extends over all active partons
($a =g,u,\bar u,\ldots,b,\bar b$), $\mu$ is the renormalization scale of the
partonic subprocesses, and $M_f$ is the fragmentation scale. 
At NLO, $M_f$ defines the point where the divergence associated with collinear
radiation off parton $a$ is to be subtracted.
This applies to the $u$, $d$, and $s$ quarks as well as for the $c$ and $b$
quarks, the masses of which are neglected.
To NLO in the modified minimal subtraction ($\overline{\rm MS}$) scheme, the 
cross sections of the relevant subprocesses are given by
\begin{eqnarray}
\label{xsection}
\frac{1}{\sigma_{tot}}\,
\frac{d\sigma_{q_i}}{dx}\left(x,\mu^2,M_f^2\right)
&\n=\n&\frac{e_{q_i}^2}{\sum_{i=1}^{n_f}e_{q_i}^2}
\left\{\delta(1-x)+\frac{\alpha_s(\mu^2)}{2\pi}
\left[P_{q\to q}^{V(0,T)}(x)\ln\frac{s}{M_f^2}+C_q(x)\right]\right\},
\nonumber\\
\frac{1}{\sigma_{tot}}\,\frac{d\sigma_g}{dx}\left(x,\mu^2,M_f^2\right)
&\n=\n&2\frac{\alpha_s(\mu^2)}{2\pi}
\left[P_{q\to g}^{(0,T)}(x)\ln\frac{s}{M_f^2}+C_g(x)\right].
\end{eqnarray}
Here, $e_{q_i}$ are the effective electroweak couplings of the quarks $q_i$ to
the $Z$-boson and the photon including propagator adjustments, and
$P_{a\to b}^{(0,T)}$ are the LO timelike splitting functions, which may be
found in the Appendix.
The $C$ functions are given by \cite{35}
\begin{eqnarray}
C_q(x)&\n=\n&
C_F\left\{
\left[-\frac{9}{2}+4\zeta(2)\right]\delta(1-x)
-\frac{3}{2}\left(\frac{1}{1-x}\right)_+
+2\left[\frac{\ln{(1-x)}}{1-x}\right]_+
+\frac{5}{2}-\frac{3}{2}x
\right.\nonumber\\
&\n+\n&\left.
4\frac{\ln x}{1-x}-(1+x)\left[2\ln x+\ln(1-x)\right]\right\},
\nonumber\\
C_g(x)&\n=\n&
C_F\frac{1+(1-x)^2}{x}\left[2\ln x+\ln(1-x)\right],
\end{eqnarray}
where $\zeta(2)=\pi^2/6$, $C_F=4/3$, and the plus distributions are defined in
the Appendix.
We evaluate $\alpha_s$ using the two-loop formula with $n_f=5$ flavours.
We identify $\mu=M_f=\sqrt s$, so that, in (\ref{xsection}),
the terms involving $\ln(s/M_f^2)$ vanish.

Up to this point, the formalism is identical to the one for describing the
fragmentation into light hadrons based on the $\overline{\rm MS}$ 
factorization scheme. 
For the fragmentation into heavy particles, we take one further step. 
As in the previous work \cite{3}, we adjust the factorization of the
final-state collinear singularities associated with the charm and bottom 
quarks in such a way that it matches the massive calculation. 
We achieve this by changing the factorization scheme. 
Specifically, we substitute in the hard-scattering cross sections 
(\ref{xsection})
$\alpha_s(\mu^2)P_{a\to Q}^{V(0,T)}(x)\ln(s/M_f^2)\rightarrow
\alpha_s(\mu^2)P_{a\to Q}^{V(0,T)}(x)\ln(s/M_f^2)
-\alpha_s\left(\mu_0^2\right)d_{Qa}(x)$,
where
\begin{eqnarray}
\label{dij}
d_{QQ}(x)
&\n=\n&-P_{Q\to Q}^{V(0,T)}(x)\ln\frac{\mu_0^2}{m_Q^2}
+C_F\left\{-2\delta(1-x)+2\left(\frac{1}{1-x}\right)_+
+4\left[\frac{\ln(1-x)}{1-x}\right]_+\right.
\nonumber\\
&\n-\n&\left.(1+x)\left[1+2\ln(1-x)\right]\right\},
\nonumber\\
d_{Qg}(x)
&\n=\n&-P_{g\to Q}^{(0,T)}(x)\ln\frac{\mu_0^2}{m_Q^2},
\nonumber\\
d_{Qq}(x)
&\n=\n&d_{Q\bar q}(x)=d_{Q\bar Q}(x)=0,
\end{eqnarray}
with $Q=c,b$ and $q=u,d,s$.
The $d_{\bar Qa}$ functions emerge by charge conjugation.
These functions are extracted from \cite{21}.
Note that $d_{Qg}(x)$ and $d_{\bar Qg}(x)$ do not yet enter (\ref{xsection})
at NLO.
The same substitutions must also be performed in any massless hard-scattering
cross sections that are to be convoluted with our FF's, in particular with
those of inclusive particle production in $ep$ collisions, which will be 
studied in the next section.
This change of factorization must be regarded in connection with our
assumptions concerning the FF's at the input scale, which refer to massive
charm and bottom quarks.
This means that, without this change, we would proceed as in the case of
fragmentation into light hadrons and obtain somewhat different parameters for
the input distributions.

The FF's in (\ref{sigx}) are parameterized at the starting scale 
$M_f=\mu_0$ by the ans\"atze (\ref{peterson}) and/or (\ref{standard}), 
depending on whether set M or S is considered.
Once we know the FF's at the scale $\mu_0$, the $\mu^2$ evolution is ruled by
the Altarelli-Parisi (AP) equations \cite{22}.
Our task is thus to determine the parameters of the distributions
(\ref{standard}) and (\ref{peterson}) at the starting scale $\mu_0$ in such a
way that the evolved FF's, when convoluted with the hard-scattering
cross sections, fit the ALEPH \cite{16} and OPAL \cite{17} data at
$M_f=M_Z$.
In the case of form~(\ref{standard}), the AP equations can be easily solved
with the help of the Mellin-transform technique, since the moments of the
starting distribution~(\ref{standard}) are simple. 
Further details are given, for example, in our earlier work~\cite{23}. 
In the case of the Peterson form~(\ref{peterson}), the computation of the
moments in Mellin space is numerically rather complicated; see \cite{3}
for the corresponding formulae.
In this case, we find it more convenient to perform the evolution of the FF's
in $x$ space by iteratively solving the AP equations in their integral form, 
\begin{equation}
\label{ap}
D_a(x,t)=D_a(x,0)+\int_0^tdt^\prime\sum_b\int_x^1\frac{dz}{z}
P_{a\to b}^{(T)}\left(\frac{x}{z},\alpha_s(t^\prime)\right)D_b(z,t^\prime),
\end{equation}
where $t=\ln\left(\mu^2/\mu_0^2\right)$.
The first term on the right-hand side of (\ref{ap}) is the distribution at
the starting scale $\mu_0$. 
As a check for the numerical stability of our procedure, we compared both
methods of evolution at LO and NLO for forms (\ref{standard}) and
(\ref{peterson}), and found good agreement.
The NLO timelike splitting functions $P_{a\to b}^{(1,T)}(x)$, which enter
(\ref{ap}), may be extracted from the literature \cite{24}.
Since, to our knowledge, there is no reference where all these functions are 
presented in a convenient form, i.e., with the coefficients of the delta
functions and plus distributions explicitly displayed, we collect them in the
Appendix.

In the case of set S, we need to fit $N$, $\alpha$, and $\beta$ in
(\ref{standard}) for the charm- and bottom-quark FF's, while, in the case
of set M, we need to determine these parameters for the bottom-quark FF as
well as $N$ and $\epsilon_c$ in (\ref{peterson}) for the charm-quark FF.
Both ALEPH \cite{16} and OPAL \cite{17} extracted the total $D^{*\pm}$
momentum distribution as well as the individual contributions due to the
$Z\to c\bar c$ and $Z\to b\bar b$ channels.
In the case of ALEPH, we received the individual $c\bar c$ and $b\bar b$ data
sets in numerical form via private communication \cite{25}.
In the ALEPH analysis, the branching ratio of $D^0 \rightarrow K^-\pi^+$
(or the charge-conjugated final state) was taken to be 0.0371.
Since this value has been superseded by the value 0.041 \cite{32}, which was
also used in the OPAL analysis, we rescale the ALEPH data with the factor
0.9049.
In our LO (NLO) analysis, we use the one-loop (two-loop) formula for
$\alpha_s$ with $\Lambda_{\overline{\rm MS}}^{(5)}=108$~MeV (227~MeV) as in
our previous work on inclusive pion and kaon production \cite{23}.
At first sight, the value $\Lambda_{\overline{\rm MS}}^{(5)}=108$~MeV of our
LO fit may seem unrealistically low; this is, however, inconsequential because
a higher value can easily be accommodated by slightly modifying the other fit
parameters without damaging the quality of the fit.
The starting scales are chosen to be $\mu_0=2m_c$ with $m_c=1.5$~GeV for
all FF's except for the case of bottom, where it is taken to be
$\mu_0=2m_b$ with $m_b=5$~GeV.

The values of $N$, $\alpha$, $\beta$, and $\epsilon_c$ resulting from the
combined LO and NLO fits to these data are displayed in Table~\ref{pars}.
The parameter $\epsilon_c$ appears only in the case of the charm-quark FF's of
set M.
In the case of the charm-quark FF's of set S, we have $\alpha\gg\beta$,
as is expected for a distribution peaked at large $x$.
On the other hand, $\beta>\alpha$ for the bottom-quark FF's, which have their
maxima at smaller values of $x$.
The parameters of the bottom-quark FF's of sets S and M are very similar,
i.e., the data fix the charm- and bottom-quark FF's independently of
each other.

In Figs.~\ref{aleph}a and b, we compare the ALEPH data \cite{16,25} with our
fits according to sets S and M, respectively.
In Figs.~\ref{opal}a and b, we do the same for the OPAL data \cite{17}.
In each case, we show both LO and NLO results.
Except for very small $x$, the LO and NLO curves are very similar. 
At small $x$, we observe large differences between LO and NLO, indicating
that, in this region, the perturbative treatment ceases to be valid. 
In this region, also the massless approximation is not valid anymore.
This may be understood by regarding the phase-space boundary for the
production of $D^{*\pm}$ mesons.
Since $D^{*\pm}$ mesons have mass $m(D^{*\pm})=2.11$~GeV \cite{32a}, they can
only be produced for $x>x_{\rm min}=2m(D^{*\pm})/\sqrt s$.
At $\sqrt s=M_Z$, one has $x_{\rm min}=0.046$.
This is just the region where our NLO results turn negative.
Therefore, our results are meaningful only for $x\simgt x_{\rm cut}=0.1$.
Consequently, we exclude the first bin of the OPAL data ($0<x<0.1$) from our
analysis.
Note that this phase-space boundary is not only theoretically motivated, but
does also turn up in the measurements, which give values consistent with zero
whenever $x<x_{\rm min}$.
This is not only true for the OPAL data, but also for data taken at lower
energies, e.g., those from ARGUS \cite{19}.

In Figs.~\ref{aleph}a, b and \ref{opal}a, b, we have plotted the fully
normalized cross section of (\ref{sigx}), which also includes the gluon
contributions in NLO.
The shown charm and bottom contributions include terms with $a=c,\bar c$ and
$a=b,\bar b$ in (\ref{sigx}), respectively, as well as the term with $a=g$
distributed to the charm and bottom terms in proportion to
$e_{q_i}^2/\sum_{i=1}^{n_f}e_{q_i}^2$, which is motivated by
(\ref{xsection}).
The contributions from $a=u,\bar u,d,\bar d,s,\bar s$ are negligible in size;
they add up to less than 2\% of the total integrated cross section and
mostly contribute at small $x$. 
This procedure approximately produces the quantities that should be compared
with the ALEPH and OPAL data.
For the time being, we ignore the effect of electromagnetic initial-state
radiation, which has not been corrected for in the data.

The quality of the fit is measured in terms of the average $\chi^2$ for all
selected data points. 
In Table~\ref{fit}, we list the $\chi^2$ values obtained with sets S and M at
LO and NLO for the ALEPH and OPAL data, and their combination.
We also specify the individual results for the $Z\to c\bar c$ and
$Z\to b\bar b$ samples of ALEPH and OPAL.
We observe that the $Z\to c\bar c$ samples yield worse $\chi^2$ values than
the $Z\to b\bar b$ samples.
The OPAL data tend to give better $\chi^2$ values than the ALEPH data.
Especially in the case of ALEPH, set S leads to better $\chi^2$ values than
set M, which is not surprising if one recalls that set S has six degrees of
freedom while set M has only five.

The integrals of the charm- and bottom-quark FF's into $D^{*\pm}$ mesons over
$x$ give the branching fractions of these transitions. 
For the reasons given above, we restrict our considerations to the region
$x_{\rm cut}<x<1$, with $x_{\rm cut}=0.1$, and define
\begin{equation}
\label{branchcut}
B_Q(\mu)=\int_{x_{\rm cut}}^1dxD_Q(x,\mu^2),
\end{equation}
where $Q=c,b$.
Experimentally, the contribution from the omitted region $0<x<x_{\rm cut}$ is
close to zero with a large error.
It is interesting to study $B_Q(\mu)$ for $Q=c,b$ at threshold $\mu=2m_Q$ and
at the $Z$-boson resonance $\mu=M_Z$.
In Table~\ref{branching}, we present the results for sets S and M at LO and 
NLO.
We observe that these branching fractions are changed very little through the
evolution from $2m_Q$ to $M_Z$, and they are almost equal for charm- and 
bottom-quark fragmentation. 
The branching fractions at the $Z$-boson resonance can be compared with the
experimental numbers reported by the OPAL collaboration \cite{17}, which
are $B_c(M_Z)=0.259$ and $B_b(M_Z)=0.252$, with errors of approximately
25\%. 
These results, which are obtained from independent fits to all experimental
data, are consistent with our fits.
At LO (NLO), the charm to bottom ratio of the branching fractions at $\mu=M_Z$
is 0.92 (0.94) for set S and 1.07 (1.01) for set M.
This agrees well with the experimental results
$B_c(M_Z)/B_b(M_Z)=1.15{+0.20\atop-0.17}({\rm stat.})$ \cite{16}
and $1.03\pm0.11({\rm stat.})\pm0.10({\rm syst.})$ \cite{17}.

Another quantity of interest is the mean momentum fraction,
\begin{equation}
\label{average}
\langle x\rangle_Q(\mu)=\frac{1}{B_Q(\mu)}\int_{x_{\rm cut}}^1dx\,
xD_Q(x,\mu^2),
\end{equation}
where $Q=c,b$.
In Table~\ref{xav}, we collect the values of $\langle x\rangle_Q(\mu)$ for
$Q=c,b$ evaluated with sets S and M in LO and NLO at $\mu=2m_Q,M_Z$.
The differences between sets S and M and between LO and NLO are minor.
However, the effect of the $\mu^2$ evolution is significant, e.g.,
$\langle x\rangle_c(\mu)$ moves from approximately 0.7 at $\mu=2m_c$ down to
about 0.5 at $\mu=M_Z$.
Our results for $\langle x\rangle_c(M_Z)$ should be compared with the 
experimental numbers reported by ALEPH \cite{16} and OPAL \cite{17},
\begin{eqnarray}
\langle x\rangle_c(M_Z)&\n=\n&0.495{+0.010\atop-0.011}({\rm stat.})
\pm0.007({\rm syst.}),
\nonumber\\
\langle x\rangle_c(M_Z)&\n=\n&0.515{+0.008\atop-0.005}({\rm stat.})
\pm0.010({\rm syst.}),
\end{eqnarray}
respectively.
We conclude that our fits are in reasonable agreement with the independent
determinations in \cite{16,17}.

Next, we test whether the FF's constructed from the ALEPH and OPAL data,
after evolution to lower scales, lead to predictions for $D^{*\pm}$ production
by $e^+e^-$ annihilation that are consistent with experimental data taken at
lower CM energies.
A number of measurements at lower energies were performed, e.g., at
DORIS, CESR, PEP, and PETRA; for a compilation, see \cite{18}.
These data have much lower statistics than the data from LEP1. 
For the comparison, we selected the data from ARGUS \cite{19} at
$\sqrt s=10.49$~GeV, from HRS \cite{27} at $\sqrt s=29$~GeV, and from
TASSO \cite{30} at $\sqrt s=34.2$~GeV. 
The TASSO collaboration measured two decay channels of the $D^0$ meson, 
namely, $D^0\rightarrow K^-\pi^+\pi^+\pi^-$ and $D^0 \rightarrow K^-\pi^+$.
The latter decay channel also allowed measurements at lower values of $x$;
we shall denote this data set by TASSO~2 and the other one by TASSO~1.
For consistency, we corrected the HRS and TASSO data by updating the values of
the branching ratios of $D^{*\pm}\rightarrow D^0 \pi^+$,
$D^0\rightarrow K^-\pi^+$, and $D^0\rightarrow K^-\pi^+\pi^+\pi^-$ according 
to the 1994 tables of the Particle Data Group \cite{32}, which were also used
by OPAL \cite{17}. 
This leads to correction factors of 0.702 for TASSO~1 and 0.677 for TASSO~2.
In the publication by HRS \cite{27}, the cross sections are not given
separately for the different decay channels, so that we can only approximate
the required correction by an overall factor of 0.877.
In Fig.~\ref{comp}, we compare the ARGUS, HRS and TASSO data on the cross 
section $sd\sigma/dx$ of $e^+e^-\to D^{*\pm}+X$ with our respective LO and NLO
predictions based on set S.
The theoretical results are calculated according to (\ref{sigx}) with
$n_f=5$ quark flavours (except for the case of ARGUS, where $n_f=4$ is used),
i.e., also the contributions due to gluon and light-flavour fragmentation are
consistently included.
We only show our results for $0.1<x<0.9$.
For completeness, Fig.~\ref{comp} also includes the comparison of the OPAL 
data \cite{17} on the dimensionless cross section
$(1/\sigma_{\rm tot})d\sigma/dx$ with our LO and NLO predictions according to
set S.
As can be seen from Fig.~\ref{comp}, the agreement between our predictions and
the data is satisfactory.
This nicely demonstrates that the data indeed exhibit the scaling violation
predicted on the basis of the LO and NLO AP evolution equations for the FF's.
In fact, the change in the shape of the differential cross section 
with $\mu^2$ is mainly due to the bottom quark.
The ARGUS data \cite{19} are taken off the resonance, at $\sqrt s=10.49$~GeV,
where the bottom quark is not yet active.
The inclusion of the bottom-quark contribution leads to a softening of the
distribution, as may be seen in the case of the HRS data \cite{27} at
$\sqrt s=29$~GeV and the TASSO data \cite{30} at $\sqrt s=34.2$~GeV.
The evolution from 29~GeV to 34.2~GeV has no discernible effect.
Going to the $Z$-boson pole at $\sqrt s=91.2$~GeV further increases the
relative importance of the bottom-quark fragmentation and leads to an even
softer spectrum.
For $0.2<x<0.8$, there is little difference between the LO and NLO cross
sections.

The successful comparisons in Fig.~\ref{comp} make us confident that our
$D^{*\pm}$ FF's, although constructed at $\sqrt s=M_Z$, also lead to useful 
descriptions of $D^{*\pm}$ fragmentation at other scales.
In the next section, we shall exploit this property together with the
universality of fragmentation to make predictions for inclusive $D^{*\pm}$
photoproduction at HERA.

\section{\boldmath{$D^{*\pm}$} Production in Low-\boldmath{$Q^2$} 
\boldmath{$ep$} Collisions}

In this section, we compare our NLO predictions for the cross section
of inclusive $D^{*\pm}$ photoproduction in $ep$ scattering at HERA with data
from the ZEUS collaboration \cite{2}. 
The calculation of this cross section proceeds as in the previous work
\cite{3}, except that we now use the FF's constructed in the last section.
For the calculation of the cross section $d^2\sigma/dy_{\rm lab}\,dp_T$,
we adopt the present HERA conditions, where $E_p=820$~GeV protons collide with
$E_e= 27.5$~GeV positrons in the laboratory frame.
We take the rapidity to be positive in the proton flight direction.
The quasi-real photon spectrum is described in the Weizs\"acker-Williams
approximation by the formula given in \cite{3}. 
This spectrum depends on the photon-energy fraction, $x=E_\gamma/E_e$, and the
maximum photon-virtuality, $Q_{\rm max}^2$.
In the ZEUS experiment \cite{2}, where the final-state electron is not
detected, $Q_{\rm max}^2=4$~GeV$^2$ and $0.147<x<0.869$, which corresponds to
$\gamma p$ CM energies of 115~GeV${}<W<280$~GeV.
We adopt these kinematic conditions in our analysis.
We work at NLO in the $\overline{\rm MS}$ scheme with $n_f=4$ flavours.
For the proton and photon PDF's we use set CTEQ4M \cite{33} with 
$\Lambda_{\overline{\rm MS}}^{(4)}=296$~MeV and set GRV~HO \cite{11} converted
to the $\overline{\rm MS}$ factorization scheme, respectively.
We evaluate $\alpha_s(\mu^2)$ from the two-loop formula with this value of
$\Lambda_{\overline{\rm MS}}^{(4)}$.
The $\Lambda_{\overline{\rm MS}}^{(4)}$ values implemented in the $D^{*\pm}$
FF's and photon PDF's are 352~MeV, which corresponds to the value
$\Lambda_{\overline{\rm MS}}^{(5)}=227$~MeV quoted in the previous section,
and 200~MeV, respectively.
We identify the factorization scales associated with the proton, photon, and
$D^{*\pm}$ meson, and collectively denote them by $M_f$. 
We choose the renormalization and factorization scales to be
$\mu=m_T$ and $M_f=2m_T$, respectively, where $m_T=\sqrt{p_T^2+m_c^2}$ is the
$D^{*\pm}$ transverse mass.
Whenever we present LO results, these are consistently computed using set
CTEQ4L \cite{33} of proton PDF's, set GRV~LO \cite{11} of photon PDF's, the
LO versions of our sets S and M of $D^{*\pm}$ FF's, the one-loop formula
for $\alpha_s$ with $\Lambda_{\overline{\rm MS}}^{(4)}=236$~MeV, and the LO
hard-scattering cross section.

The photoproduction cross section is a superposition of the direct- and
resolved-photon contributions.
In our NLO analysis, the resolved-photon contribution is larger than the
direct one for moderate $p_T$. 
This statement depends, however, on the factorization scheme;
only the sum of both contributions is a physical observable and can be
compared with experimental data. 
The bulk of the resolved-photon cross section is due to the charm component of
the photon PDF \cite{3}.

We first consider the $p_T$ distribution $d\sigma/dp_T$ integrated over the
rapidity interval $-1.5<y_{\rm lab}<1$ as in the ZEUS analysis \cite{2}. 
In Fig.~\ref{zeus}a, we compare the respective ZEUS data \cite{2} with our LO 
and NLO predictions based on sets S and M.
The theoretical results are multiplied with the factor 1.06 to account for the
fact that OPAL \cite{17} used the $D^{*\pm}$ and $D^0$ branching fractions of
\cite{32}, while ZEUS \cite{2} used those of \cite{32a}.
The NLO distributions fall off slightly less strongly with increasing $p_T$
than the LO ones.
It is remarkable that there is hardly any difference between the set-S and
set-M results; in fact, the LO-S and LO-M (NLO-S and NLO-M) curves almost
coincide.
This means that the details of charm-quark fragmentation are very tightly
constrained by the LEP1 data, and that the considered variation in the
functional form of the charm-quark FF at the starting scale has very little
influence on the $p_T$ distribution. 
The NLO curves should be compared with Fig.~8c in \cite{3}, where the
charm-quark FF at the starting scale $\mu_0=2m_c=3$~GeV was taken to be of
the Peterson form with $\epsilon_c=0.06$ and $B_c(2m_c)=0.260$.
This corresponds to $N=0.163$.
We recall that, in the case of set NLO~M, we have $\epsilon_c=0.0204$,
$N=0.0677$, and $B_c(2m_c)=0.232$; see Tables~\ref{pars} and \ref{branching}.
We observe that the result for $\mu=M_f/2=m_T$ in Fig.~8c of \cite{3} is 
somewhat smaller than the NLO results of Fig.~\ref{zeus}a, but has a similar 
shape.
The agreement with the data is good, even at small $p_T$, where the massless
approach ceases to be valid.

In Fig.~\ref{zeus}b, we compare the $y_{\rm lab}$ line shape of
$d^2\sigma/dy_{\rm lab}\,dp_T$, integrated over the $p_T$ interval 
3~GeV${}<p_T<12$~GeV, with the respective ZEUS data points \cite{2}.
As in Fig.~\ref{zeus}a, we consider the four cases LO~S, NLO~S, LO~M, and
NLO~M.
Again, the LO-S (NLO-S) results almost perfectly coincide with the LO-M
(NLO-M) results, i.e., the functional form of the charm-quark FF at the
starting scale is of minor significance.
The agreement between theory and experiment is reasonable.
If we compare the NLO results of Fig.~\ref{zeus}b with the result for
$\mu=M_f/2=m_T$ in Fig.~9c of \cite{3}, we see that the latter is 
somewhat smaller, but has essentially the same shape.

In \cite{2}, the cross section $d^2\sigma/dy_{\rm lab}\,dp_T$ was also
measured at three different values of $W$.
To this end, $y_{\rm lab}$ and $p_T$ were sampled in the intervals
$-1.5<y_{\rm lab}<1$ and 3~GeV${}<p_T<12$~GeV, respectively.
In Fig.~\ref{zeus}c, we show the $W$ dependence of
$d^2\sigma/dy_{\rm lab}\,dp_T$,
integrated over these $y_{\rm lab}$ and $p_T$ intervals, and compare it with
the corresponding ZEUS data \cite{2}.
Similarly to Figs.~\ref{zeus}a and b, we consider sets S and M at LO and NLO.
Within the rather large experimental errors, all four predictions agree
reasonably well with the data. 
In \cite{3}, such a figure was not presented, but it can be found in the ZEUS
publication \cite{2}.
Again, we observe that the input assumptions, encoded in sets S and M, have
practically no influence on the result.

When we fitted to the LEP1 data, we also had to take into account $D^{*\pm}$
production by bottom-quark fragmentation, as it gives a sizeable contribution
at the $Z$-boson pole.
In $ep$  collisions, however, the bottom quark does not contribute at all to
inclusive $D^{*\pm}$ photoproduction below the threshold at $m_T=m_b$.
In contrast to $e^+e^-$ annihilation, the bottom quark does not contribute
significantly even above the threshold.
To elucidate this point, we repeat the analysis of Fig.~\ref{zeus}a for
$n_f=5$ with and without the bottom-quark FF and plot the ratio in
Fig.~\ref{ratio}.
We see that, in all four cases, the relative contributions due to the
bottom-quark FF's are below 5\% and decrease with increasing $p_T$.
This may be understood by observing that, in the case of $D^{*\pm}$ 
photoproduction in $ep$ collisions, the main contribution to the cross section
arises at large $x$, typically at $x\simgt0.5$, where
$D_b(x,M_f^2)\ll D_c(x,M_f^2)$ because the bottom-quark FF's have a soft
spectrum.
For the same reason, also the contributions of the $u$, $d$, and $s$ quarks
are negligible.

We have seen that our theoretical predictions for $D^{*\pm}$ photoproduction
at HERA are very insensitive to the specific form of the input distribution of 
the charm-quark FF, and that the influence of the other quark flavours is
marginal.
When the FF templates of sets S or M are fitted to the most accurate data, we
obtain an almost unique prediction for the charm-quark FF.
In fact, the largest remaining uncertainty in the calculation of the
$D^{*\pm}$ photoproduction cross section is related to the charm content of
the resolved photon.
This offers us the possibility to employ our well constrained FF's to extract
information about the charm distribution in the photon.
It is well known that the most sensitive distributions to study such
dependences are those differential in rapidity.
This fact has been previously exploited, e.g., in \cite{34},
where we emphasized the potential to determine the gluon PDF of the photon
by accurately measuring the inclusive charged-particle rapidity spectrum in
photoproduction.
To illustrate the sensitivity to the charm PDF of the photon, we repeat the
NLO-S analysis of Fig.~\ref{zeus}b with this PDF switched off, and compare the
outcome with the full calculation in Fig.~\ref{charm}.
For comparison, also the direct-photon contribution is shown.
We observe that 73\% to 100\% (50\% to 89\%) of the resolved-photon (full)
cross section is induced by the charm content of the photon, which is most 
important in the backward direction.

Inspired by this observation, we now repeat the NLO-S analysis of
Fig.~\ref{zeus}b using in turn the NLO photon PDF sets of \cite{12,13},
which we denote by GS~HO and ACFGP, respectively.
The authors of \cite{13} considered both massless and massive charm quarks.
In their massive-charm set, the charm-quark PDF is implemented with the
boundary condition of \cite{13a} at threshold.
We observe a significant variation of the cross section, especially
in the backward region, where our predictions for set GRV~HO \cite{11}
and ACFGP (massive $c$) \cite{13} differ by more than a factor of two!
Unfortunately, the data are not yet accurate enough to rule out any of the 
available sets.
Of course, in order to reliably constrain the charm PDF of the photon, it 
would be preferable to have data with an increased low-$p_T$ cut, since our 
massless charm-quark theory is more reliable in the large-$p_T$ region.
However, this is a very interesting measurement that has a vast potential
for improving our knowledge about the charm PDF of the photon.

\section{Conclusions}

In this paper, we determined new sets of $D^{*\pm}$ FF's at LO and NLO from
fits to precise cross-section data obtained in $e^+e^-$ annihilation
by ALEPH and OPAL at LEP1.
Our analysis was based on the massless-charm factorization approach recently 
developed in \cite{3}.
In these LEP1 data, the fragmentation of charm and bottom quarks into
$D^{*\pm}$ mesons was disentangled, so that the respective FF's are separately
constrained.
For the description at the charm-quark FF, we used two different ans\"atze at 
the starting scale, namely, the three-parameter standard form conventionally
used for light-meson production and the two-parameter Peterson form.
It turned out that both prescriptions lead to excellent fits of the LEP1 data,
and that the charm- and bottom-quark FF's are each tightly constrained.
With this method, the fragmentation at any other scale is predictable via the
AP evolution, which we performed both at LO and NLO.
This enabled us to check our FF's against $e^+e^-$ data at lower CM energies.
In particular, we found very good agreement with data from TASSO \cite{30},
HRS \cite{27}, and ARGUS \cite{19}, which demonstrates that the scaling 
violation encoded in the AP equations is consistent with the data.
We stress that the results for the starting parameters are correlated with the
starting scale.
In the case of the Peterson description for charm-quark fragmentation at LO
(NLO), the epsilon parameter and $c\to D^{*+}$ branching fraction at the
starting scale $\mu_0=2m_c$ were found to be $\epsilon_c=0.0856$ (0.0204) and
$B_c(2m_c)=0.246$ (0.232), respectively.
This corresponds to an $x$ mean value of $\langle x\rangle_c(2m_c)=0.644$
(0.754).
While $B_c(\mu)$ is relatively stable under evolution to $\mu=M_Z$,
$\langle x\rangle_c(\mu)$ is appreciably lowered, down to
$\langle x\rangle_c(M_Z)=0.490$ (0.556).

We emphasize that the $D^{*\pm}$ FF's thus obtained are universal and may be
used to make predictions for inclusive $D^{*\pm}$ production in any other
high-energy experiment.
We exploited this fact by calculating the cross section of inclusive
$D^{*\pm}$ photoproduction at HERA, which we compared with the latest data
from the ZEUS collaboration.
We found that the bottom-quark contribution is marginal, and that the results
do not depend on whether we use the standard or Peterson-type charm-quark
FF's, which shows that we have good control over the fragmentation aspect of
$D^{*\pm}$ photoproduction.
Under the condition that more precise experimental data, in particular in the 
large-$p_T$ region, will be collected at HERA, this will allow one to 
constrain other non-perturbative input information such as the charm-quark PDF
of the photon, which is still poorly known.

\bigskip
\centerline{\bf NOTE ADDED IN PROOFS}
\smallskip\noindent
After the completion of our manuscript, a related paper \cite{36} has
appeared.
The authors of \cite{36} determine $D^{*\pm}$ FF's by separately fitting
$e^+e^-$ data from DORIS and LEP1.
In contrast to our procedure, they also include the $d_{Qa}$ functions of
(\ref{dij}) in the evolution, and treat the bottom quark on the same footing 
as the $u$, $d$, and $s$ quarks, i.e., they only take the charm-quark FF to be
non-vanishing at the starting scale.

\bigskip
\centerline{\bf ACKNOWLEDGMENTS}
\smallskip\noindent
We are grateful to Paul Colas for making available to us the ALEPH data
on the charm and bottom contributions to $D^{*\pm}$ production \cite{25},
and to Carsten Coldewey for providing us with the ZEUS data on $D^{*\pm}$
photoproduction \cite{2} prior to their official publication.
One of us (GK) thanks the Theory Group of the Werner-Heisenberg-Institut for
the hospitality extended to him during a visit when this paper was finalized.

\begin{appendix}

\section{Appendix: NLO timelike splitting functions}

Although the NLO timelike splitting functions have been calculated by various 
groups \cite{24}, we are not aware of any paper where they are presented in a 
way ready to use, i.e., with the coefficients of $\delta(1-x)$ and
$1/(1-x)_+$ displayed explicitly.
For the reader's convenience, we shall do this here.

Let $D_g(x,\mu^2)$, $D_{q_i}(x,\mu^2)$, and $D_{\bar q_i}(x,\mu^2)$ be the
FF's of the gluon $g$ and $n_f$ quarks $q_i$ and antiquarks $\bar q_i$
($i=1,\ldots,n_f$), respectively, into some hadron with momentum fraction $x$
at fragmentation scale $\mu$.
The $\mu^2$ evolution of these FF's is conveniently formulated for the
combinations
\begin{eqnarray}
D_i^{(\pm)}(x,\mu^2)&\n=\n&D_{q_i}(x,\mu^2)\pm D_{\bar q_i}(x,\mu^2),
\nonumber\\
D^{(\pm)}(x,\mu^2)&\n=\n&\sum_{i=1}^{n_f}D_i^{(\pm)}(x,\mu^2)
\end{eqnarray}
as
\begin{eqnarray}
&\n\n\n&\frac{\mu^2d}{d\mu^2}
\left(D_i^{(+)}(x,\mu^2)-\frac{1}{n_f}D^{(+)}(x,\mu^2)\right)
=\int_x^1\frac{dz}{z}P_{(+)}^{(T)}\left(\frac{x}{z},\alpha_s(\mu^2)\right)
\\
&\n\n\n&\hspace*{7cm}\times
\left(D_i^{(+)}(z,\mu^2)-\frac{1}{n_f}D^{(+)}(z,\mu^2)\right),
\nonumber\\
&\n\n\n&\frac{\mu^2d}{d\mu^2}D_i^{(-)}(x,\mu^2)
=\int_x^1\frac{dz}{z}P_{(-)}^{(T)}\left(\frac{x}{z},\alpha_s(\mu^2)\right)
D_i^{(-)}(z,\mu^2),
\nonumber\\
&\n\n\n&\frac{\mu^2d}{d\mu^2}D^{(+)}(x,\mu^2)
=\int_x^1\frac{dz}{z}\left[
P_{q\to q}^{(T)}\left(\frac{x}{z},\alpha_s(\mu^2)\right)D^{(+)}(z,\mu^2)
+2n_fP_{q\to g}^{(T)}\left(\frac{x}{z},\alpha_s(\mu^2)\right)D_g(z,\mu^2)
\right],
\nonumber\\
&\n\n\n&\frac{\mu^2d}{d\mu^2}D_g(x,\mu^2)
=\int_x^1\frac{dz}{z}\left[\frac{1}{2n_f}
P_{g\to q}^{(T)}\left(\frac{x}{z},\alpha_s(\mu^2)\right)D^{(+)}(z,\mu^2)
+P_{g\to g}^{(T)}\left(\frac{x}{z},\alpha_s(\mu^2)\right)D_g(z,\mu^2)\right],
\nonumber
\end{eqnarray}
where $P_{a\to b}^{(T)}\Bigl(x,\alpha_s(\mu^2)\Bigr)$ are the timelike $a\to b$
splitting functions.
$P_{(\pm)}^{(T)}\Bigl(x,\alpha_s(\mu^2)\Bigr)$ and
$P_{q\to q}^{(T)}\Bigl(x,\alpha_s(\mu^2)\Bigr)$ are decomposed into valence
(non-singlet) and sea (pure-singlet) components as
\begin{eqnarray}
P_{(\pm)}^{(T)}\Bigl(x,\alpha_s(\mu^2)\Bigr)&\n=\n&
P_{q\to q}^{V(T)}\Bigl(x,\alpha_s(\mu^2)\Bigr)\pm
P_{q\to\bar q}^{V(T)}\Bigl(x,\alpha_s(\mu^2)\Bigr),
\nonumber\\
P_{q\to q}^{(T)}\Bigl(x,\alpha_s(\mu^2)\Bigr)&\n=\n&
P_{(+)}^{(T)}\Bigl(x,\alpha_s(\mu^2)\Bigr)+
2n_fP_{q\to q}^{S(T)}\Bigl(x,\alpha_s(\mu^2)\Bigr).
\end{eqnarray}
All these splitting function have perturbative expansions of the form
\begin{equation}
P_{i\to j}^{(T)}\left(x,\alpha_s(\mu^2)\right)=
\frac{\alpha_s(\mu^2)}{2\pi}P_{i\to j}^{(0,T)}(x)
+\left(\frac{\alpha_s(\mu^2)}{2\pi}\right)^2P_{i\to j}^{(1,T)}(x)
+{\cal O}\left(\alpha_s^3\right).
\end{equation}
The LO coefficients read
\begin{eqnarray}
P_{q\to q}^{V(0,T)}(x)&\n=\n&
C_F\left[\frac{3}{2}\delta(1-x)+2\left(\frac{1}{1-x}\right)_+-1-x\right],
\nonumber\\
P_{q\to\bar q}^{V(0,T)}(x)&\n=\n&P_{q\to q}^{S(0,T)}(x)=0,
\nonumber\\
P_{q\to g}^{(0,T)}(x)&\n=\n&C_Fp_{FG}(x),
\nonumber\\
P_{g\to q}^{(0,T)}(x)&\n=\n&2Tn_fp_{GF}(x),
\nonumber\\
P_{g\to g}^{(0,T)}(x)&\n=\n&
\left(\frac{11}{6}C_A-\frac{2}{3}Tn_f\right)\delta(1-x)
+2C_A\left[\left(\frac{1}{1-x}\right)_++\frac{1}{x}-2+x-x^2\right],
\end{eqnarray}
and the NLO coefficients read
\begin{eqnarray}
P_{q\to q}^{V(1,T)}(x)&\n=\n&
\left\{C_F^2\left[\frac{3}{8}-3\zeta(2)+6\zeta(3)\right]
+C_FC_A\left[\frac{17}{24}+\frac{11}{3}\zeta(2)-3\zeta(3)\right]-C_FTn_f
\right.\nonumber\\
&\n\times\n&\left.
\left[\frac{1}{6}+\frac{4}{3}\zeta(2)\right]\right\}\delta(1-x)
+\left\{C_FC_A\left[\frac{67}{9}-2\zeta(2)\right]
-\frac{20}{9}C_FTn_f\right\}\left(\frac{1}{1-x}\right)_+
\nonumber\\ &\n+\n&
C_F^2\left\{-5(1-x)-\frac{1}{2}(7+3x)\ln x+\frac{1}{2}(1+x)\ln^2x
+p_{FF}(x)\ln x
\right.\nonumber\\ &\n\times\n&\left.
\left[\frac{3}{2}-2\ln x+2\ln(1-x)\right]\right\}
+C_FC_A\left\{\frac{1}{18}(53-187x)+(1+x)[\zeta(2)+\ln x]
\right.\nonumber\\ &\n+\n&\left.
\frac{1}{2}p_{FF}(x)\ln x\left(\frac{11}{3}+\ln x\right)\right\}
+C_FTn_f\left[\frac{2}{9}(-1+11x)-\frac{2}{3}p_{FF}(x)\ln x\right],
\nonumber\\
P_{q\to\bar q}^{V(1,T)}(x)&\n=\n&
C_F\left(2C_F-C_A\right)\left[2(1-x)+(1+x)\ln x+p_{FF}(-x)S_2(x)
\right],
\nonumber\\
P_{q\to q}^{S(1,T)}(x)&\n=\n&
C_FT\left[4\left(-\frac{5}{9x}-2+x+\frac{14}{9}x^2\right)
-\left(5+9x+\frac{8}{3}x^2\right)\ln x+(1+x)\ln^2x\right],
\nonumber\\
P_{q\to g}^{(1,T)}(x)&\n=\n&
C_F^2\left\{\frac{1}{2}(-1+9x)+\left(-8+\frac{x}{2}\right)\ln x
+2x\ln(1-x)+\left(1-\frac{x}{2}\right)\ln^2x+p_{FG}(x)
\right.\nonumber\\ &\n\times\n&\left.
\left[-8\zeta(2)+4\ln x\ln(1-x)+\ln^2(1-x)-8S_1(x)\right]\right\}
+C_FC_A\left\{\frac{1}{9}\left(62-\frac{35}{2}x
\right.\right.\nonumber\\ &\n-\n&\left.
44x^2\right)+2\left(1+6x+\frac{4}{3}x^2\right)\ln x-2x\ln(1-x)-(4+x)\ln^2x
+p_{FG}(x)
\nonumber\\ &\n\times\n&
\left[\frac{17}{18}+7\zeta(2)-3\ln x-\frac{3}{2}\ln^2x-2\ln x\ln(1-x)
-\ln^2(1-x)+8S_1(x)\right]
\nonumber\\ &\n+\n&\left.
p_{FG}(-x)S_2(x)\right\},
\nonumber\\
P_{g\to q}^{(1,T)}(x)&\n=\n&
C_FTn_f\left\{-2+3x+(-7+8x)\ln x-4\ln(1-x)+(1-2x)\ln^2x
+2p_{GF}(x)
\right.\nonumber\\ &\n\times\n&
\left[-5+6\zeta(2)+\ln x-\ln(1-x)-\ln^2x-2\ln x\ln(1-x)-\ln^2(1-x)
\right.\nonumber\\ &\n+\n&\left.\left.
8S_1(x)\right]\right\}
+C_ATn_f\left\{\frac{2}{9}\left(-\frac{20}{x}-76+83x\right)
-\frac{4}{3}(1+19x)\ln x+4\ln(1-x)
\right.\nonumber\\ &\n+\n&
2(1+4x)\ln^2x
+p_{GF}(x)\left[\frac{178}{9}-14\zeta(2)-\frac{4}{3}\ln x+\frac{10}{3}\ln(1-x)
-\ln^2x
\right.\nonumber\\ &\n+\n&\left.\left.
8\ln x\ln(1-x)+2\ln^2(1-x)-16S_1(x)\right]+2p_{GF}(-x)S_2(x)\right\}
\nonumber\\ &\n+\n&
(Tn_f)^2\left\{-\frac{8}{3}-\frac{8}{3}p_{GF}(x)\left[\frac{2}{3}+\ln x
+\ln(1-x)\right]\right\},
\nonumber\\
P_{g\to g}^{(1,T)}(x)&\n=\n&
\left[C_A^2\left(\frac{8}{3}+3\zeta(3)\right)
-C_FTn_f-\frac{4}{3}C_ATn_f\right]\delta(1-x)
+\left[C_A^2\left(\frac{67}{9}-2\zeta(2)\right)
\right.\nonumber\\ &\n-\n&\left.
\frac{20}{9}C_ATn_f\right]\left(\frac{1}{1-x}\right)_+
+C_FTn_f\left[4\left(\frac{23}{9x}-1+3x-\frac{41}{9}x^2\right)
+2\left(\frac{8}{3x}+5
\right.\right.\nonumber\\ &\n+\n&\left.\left.
7x+\frac{8}{3}x^2\right)\ln x+2(1+x)\ln^2x\right]
+C_ATn_f\left[\frac{2}{9}\left(-\frac{23}{x}+29-19x+23x^2\right)
\right.\nonumber\\ &\n-\n&\left.
\frac{4}{3}(1+x)\ln x-\frac{8}{3}p_{GG}(x)\ln x\right]
+C_A^2\left\{-\frac{1}{18}(25+109x)-2\zeta(2)
\right.\nonumber\\ &\n\times\n&\left.
\left(\frac{1}{x}-2+x-x^2\right)
+\frac{1}{3}\left(-\frac{44}{x}+11-25x\right)\ln x-4(1+x)\ln^2x
\right.\nonumber\\ &\n+\n&\left.
p_{GG}(x)\ln x\left[\frac{22}{3}-3\ln x+4\ln(1-x)\right]
+2p_{GG}(-x)S_2(x)\right\}.
\end{eqnarray}
Here, $\zeta$ is Riemann's zeta function, with values $\zeta(2)=\pi^2/6$ 
and $\zeta(3)\approx1.202$, and we have used the auxiliary functions
\begin{eqnarray}
p_{FF}(x)&\n=\n&\frac{1+x^2}{1-x},
\nonumber\\
p_{FG}(x)&\n=\n&\frac{1+(1-x)^2}{x},
\nonumber\\
p_{GF}(x)&\n=\n&x^2+(1-x)^2,
\nonumber\\
p_{GG}(x)&\n=\n&\frac{[1-x(1-x)]^2}{x(1-x)},
\end{eqnarray}
and
\begin{eqnarray}
S_1(x)&\n=\n&-\li(1-x),\nonumber\\
S_2(x)&\n=\n&-\zeta(2)+\frac{1}{2}\ln^2x-2\ln x\ln(1+x)-2\li(-x),
\end{eqnarray}
where $\li(x)=-\int_0^xdt\,\ln(1-t)/t$ is the dilogarithm.
In the case of the colour gauge group SU($N_c$), the Casimir operators of the
fundamental and adjoint representations have the eigenvalues
$C_F=\left(N_c^2-1\right)/(2N_c)$ and $C_A=N_c$, respectively.
The trace normalization of the fundamental representation is $T=1/2$.

\end{appendix}

\newpage

\begin{table}
\centerline{\bf TABLE CAPTIONS}
\bigskip

\caption{Fit parameters for the charm- and bottom-quark FF's of sets S and M
at LO and NLO.
The corresponding starting scales are $\mu_0=2m_c=3$~GeV and 
$\mu_0=2m_b=10$~GeV, respectively.
All other FF's are taken to be zero at $\mu_0=2m_c$
\protect\label{pars}}

\caption{$\chi^2$ per degree of freedom pertaining to the LO and NLO fits of
types S and M to the ALEPH \protect\cite{16} and OPAL \protect\cite{17}
samples of $Z\to c\bar c$ and $Z\to b\bar b$ and their combinations.
The first bin of the OPAL data has been excluded
\protect\label{fit}}

\caption{Branching fractions of charm and bottom quarks into $D^{*\pm}$ mesons
at the respective starting scales and at $\mu=M_Z$ evaluated from 
(\protect\ref{branchcut}) with sets S and M at LO and NLO
\protect\label{branching}}

\caption{Average momentum fractions of $D^{*\pm}$ mesons produced through 
charm- and bottom-quark fragmentation at the respective starting scales and at
$\mu=M_Z$ evaluated from (\protect\ref{average}) with sets S and M at LO
and NLO
\protect\label{xav}}

\end{table}

\newpage

\begin{figure}
\centerline{\bf FIGURE CAPTIONS}
\bigskip

\caption{The cross sections of inclusive $D^{*\pm}$ production in $e^+e^-$
annihilation evaluated with (a) set S and (b) set M at LO and NLO are compared
with the ALEPH data \protect\cite{16}.
The three sets of curves and data correspond to the $Z\to c\bar c$ and
$Z\to b\bar b$ samples as well as their combination
\protect\label{aleph}}

\caption{As Fig.~\protect\ref{aleph}, but for the OPAL data \protect\cite{17}
\protect\label{opal}\hskip8cm}

\caption{The ARGUS \protect\cite{19}, HRS \protect\cite{27}, TASSO
\protect\cite{30}, and OPAL \protect\cite{17} data on $D^{*\pm}$ production in
$e^+e^-$ annihilation are compared with the LO and NLO predictions based on
set S.
For separation, the data have been rescaled by powers of 10.
In the case of TASSO, the open triangles refer to the
$D^0\rightarrow K^-\pi^+\pi^+\pi^-$ channel (TASSO~1) and the solid triangles 
to the $D^0 \rightarrow K^-\pi^+$ channel (TASSO~2).
In the case of OPAL, we consider the dimensionless quantity
$(1/\sigma_{\rm tot})d\sigma/dx$ and discard the data point at $x=0.05$
\protect\label{comp}}

\caption{The LO and NLO predictions of inclusive $D^{*\pm}$ photoproduction
in $ep$ collisions based on sets S and M are compared with the ZEUS data
\protect\cite{2}.
We consider (a) the $p_T$ distribution $d\sigma/dp_T$ integrated over
$-1.5<y_{\rm lab}<1$ and 115~GeV${}<W<280$~GeV,
(b) the $y_{\rm lab}$ distribution $d\sigma/dy_{\rm lab}$ integrated over
3~GeV${}<p_T<12$~GeV and 115~GeV${}<W<280$~GeV, and
(c) the $W$ distribution $d\sigma/dW$ integrated over
3~GeV${}<p_T<12$~GeV and $-1.5<y_{\rm lab}<1$
\protect\label{zeus}}

\caption{Influence of the bottom-quark FF's on the LO and NLO predictions of
inclusive $D^{*\pm}$ photoproduction in $ep$ collisions based on sets S and M.
The analysis of Fig.~\ref{zeus}a is repeated for $n_f=5$ with and without the
bottom-quark FF's, and the ratio of the two results is shown as a function of 
$p_T$
\protect\label{ratio}}

\caption{Influence of the charm-quark PDF of the photon on the NLO prediction
of inclusive $D^{*\pm}$ photoproduction in $ep$ collisions based on set S.
The analysis of Fig.~\ref{zeus}b is repeated with this PDF switched off.
For comparison, also the direct-photon contribution is shown
\protect\label{charm}}

\caption{Influence of the photon PDF's on the NLO prediction of inclusive
$D^{*\pm}$ photoproduction in $ep$ collisions based on set S.
The analysis of Fig.~\ref{zeus}b is repeated using set GS~HO \protect\cite{12}
and the ACFGP \protect\cite{13} sets for massless and massive charm quarks
\protect\label{pdf}}

\end{figure}

\newpage

\begin{table}
\begin{center}
\begin{tabular}{|c|c||c|c|c|c|}
\hline
set & flavour & $N$   & $\alpha$ & $\beta$ & $\epsilon_c$ \\
\hline
\hline
LO S & $c$ & 449 & 8.01 & 3.02 & -- \\
\cline{2-6}
     & $b$ & 335 & 3.24 & 6.47 & -- \\
\hline
NLO S & $c$ & 69.7 & 8.78 & 1.60 & -- \\
\cline{2-6}
      & $b$ & 260 & 3.33 & 5.83 & -- \\
\hline
\hline
LO M & $c$ & 0.202 & -- & -- & 0.0856 \\
\cline{2-6}
     & $b$ & 235 & 3.15 & 6.06 & -- \\
\hline
NLO M & $c$ & 0.0677 & -- & -- & 0.0204 \\
\cline{2-6}
     & $b$ & 214 & 3.25 & 5.65 & -- \\
\hline 
\end{tabular}

\smallskip
{\bf Table~1}

\vspace{1cm}

\begin{tabular}{|c||c||c|c|c||c|c|c|}
\hline
set & total & \multicolumn{3}{|c||}{ALEPH}& \multicolumn{3}{|c|}{OPAL}\\
      &       & sum  & $c$  & $b$  & sum  & $c$   & $b$ \\
\hline
\hline
LO S  & 0.85 & 0.84 & 1.54 & 0.78 & 0.44 & 1.05 & 0.66 \\
\hline
NLO S & 0.90 & 0.96 & 1.61 & 0.79 & 0.52 & 1.12 & 0.55 \\
\hline
\hline
LO M  & 1.05 & 1.25 & 2.43 & 0.92 & 0.37 & 0.99 & 0.68 \\
\hline
NLO M & 0.96 & 1.21 & 2.08 & 0.82 & 0.37 & 0.94 & 0.56 \\
\hline
\end{tabular}

\smallskip
{\bf Table~2}

\vspace{1cm}

\begin{tabular}{|c||c|c|c|c|}
\hline
set & $B_c(2m_c)$ & $B_c(M_Z)$ & $B_b(2m_b)$ & $B_b(M_Z)$ \\
\hline 
\hline
LO S  & 0.220 & 0.210 & 0.242 & 0.228 \\
\hline 
NLO S & 0.217 & 0.209 & 0.236 & 0.223 \\
\hline
\hline
LO M  & 0.246 & 0.231 & 0.229 & 0.215 \\
\hline
NLO M & 0.232 & 0.221 & 0.231 & 0.219 \\
\hline
\end{tabular}

\smallskip
{\bf Table~3}

\vspace{1cm}

\begin{tabular}{|c||c|c|c|c|}
\hline
set & $\langle x\rangle_c(2m_c)$ & $\langle x\rangle_c(M_Z)$ &
$\langle x\rangle_b(2m_b)$ & $\langle x\rangle_b(M_Z)$ \\
\hline 
\hline
LO S  & 0.691 & 0.520 & 0.365 & 0.321 \\
\hline 
NLO S & 0.790 & 0.579 & 0.390 & 0.340 \\
\hline
\hline
LO M  & 0.644 & 0.490 & 0.373 & 0.328 \\
\hline
NLO M & 0.754 & 0.556 & 0.392 & 0.342 \\
\hline
\end{tabular}

\smallskip
{\bf Table~4}
\end {center}
\end{table}

\newpage
\begin{figure}[ht]
\epsfig{figure=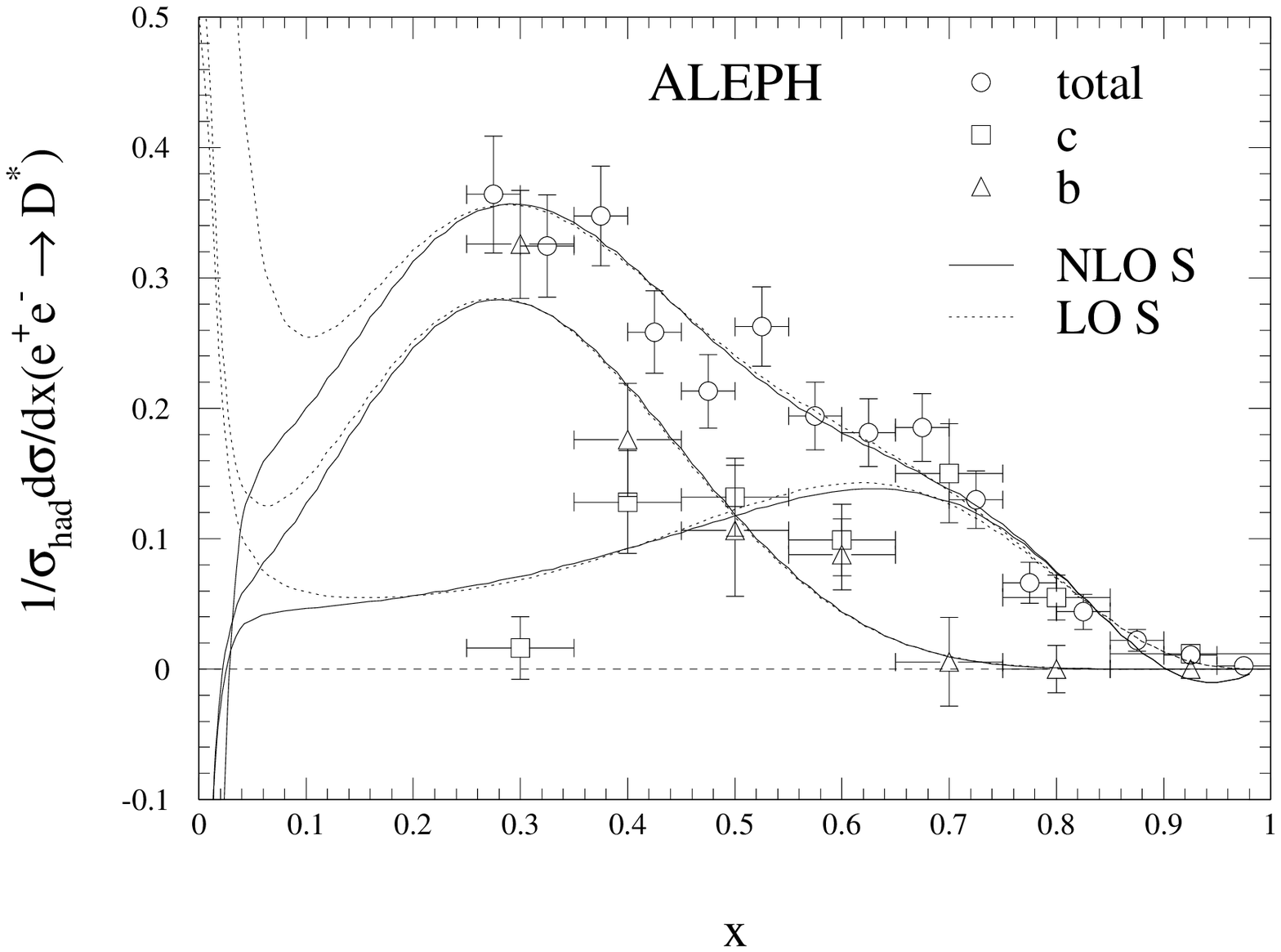,width=\textwidth}
\centerline{\Large\bf Fig.~1a}
\end{figure}

\newpage
\begin{figure}[ht]
\epsfig{figure=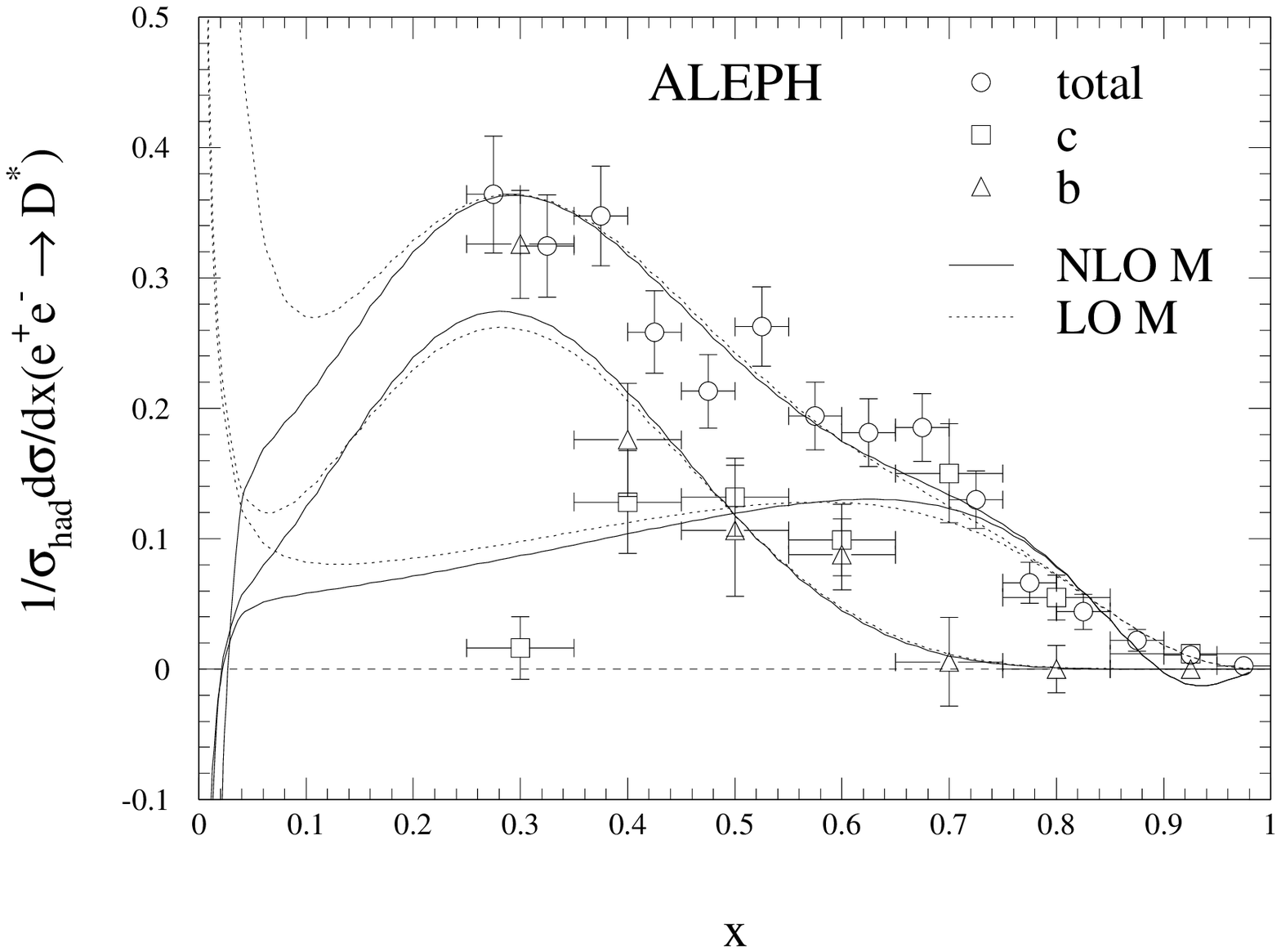,width=\textwidth}
\centerline{\Large\bf Fig.~1b}
\end{figure}

\newpage
\begin{figure}[ht]
\epsfig{figure=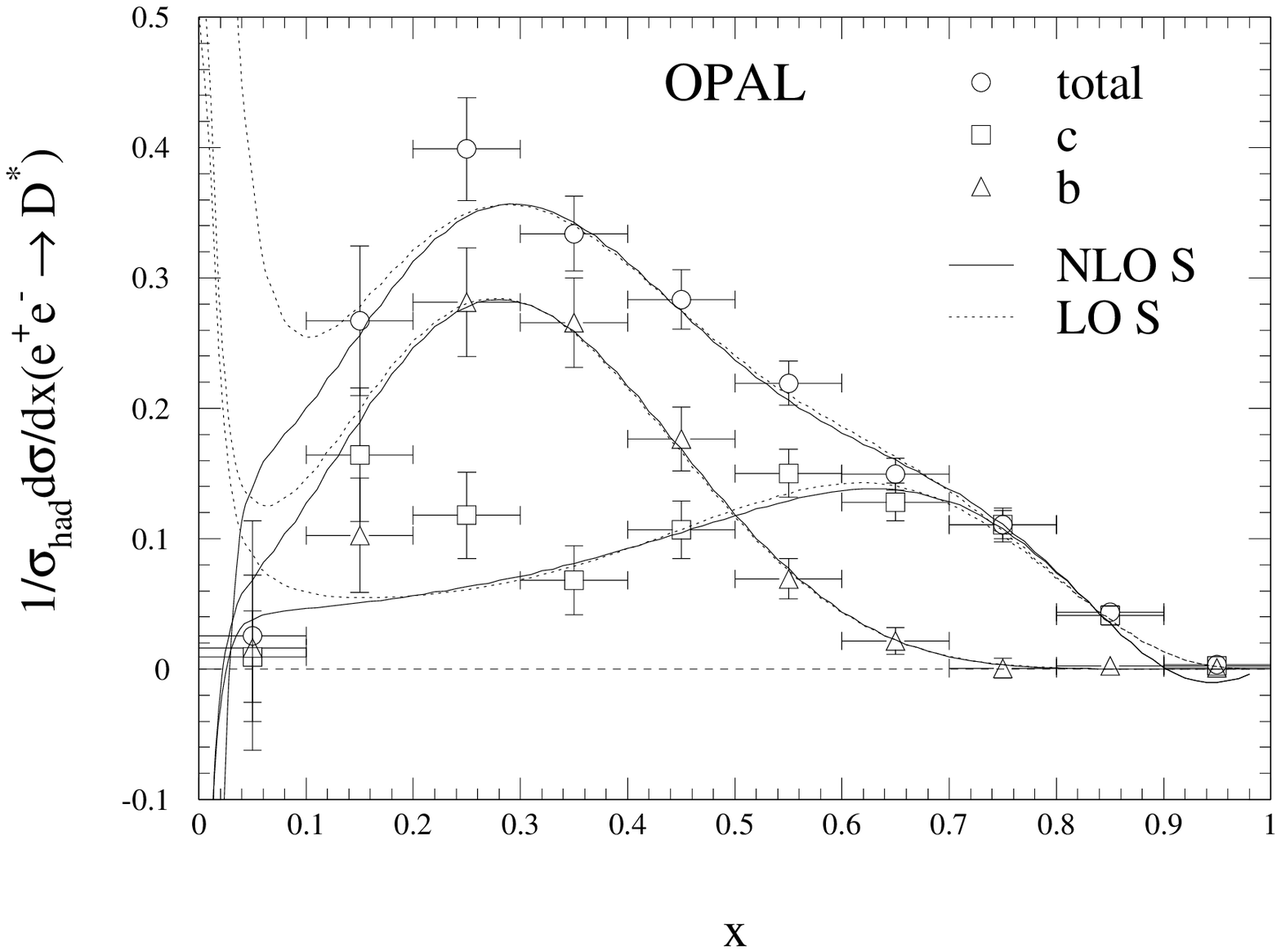,width=\textwidth}
\centerline{\Large\bf Fig.~2a}
\end{figure}

\newpage
\begin{figure}[ht]
\epsfig{figure=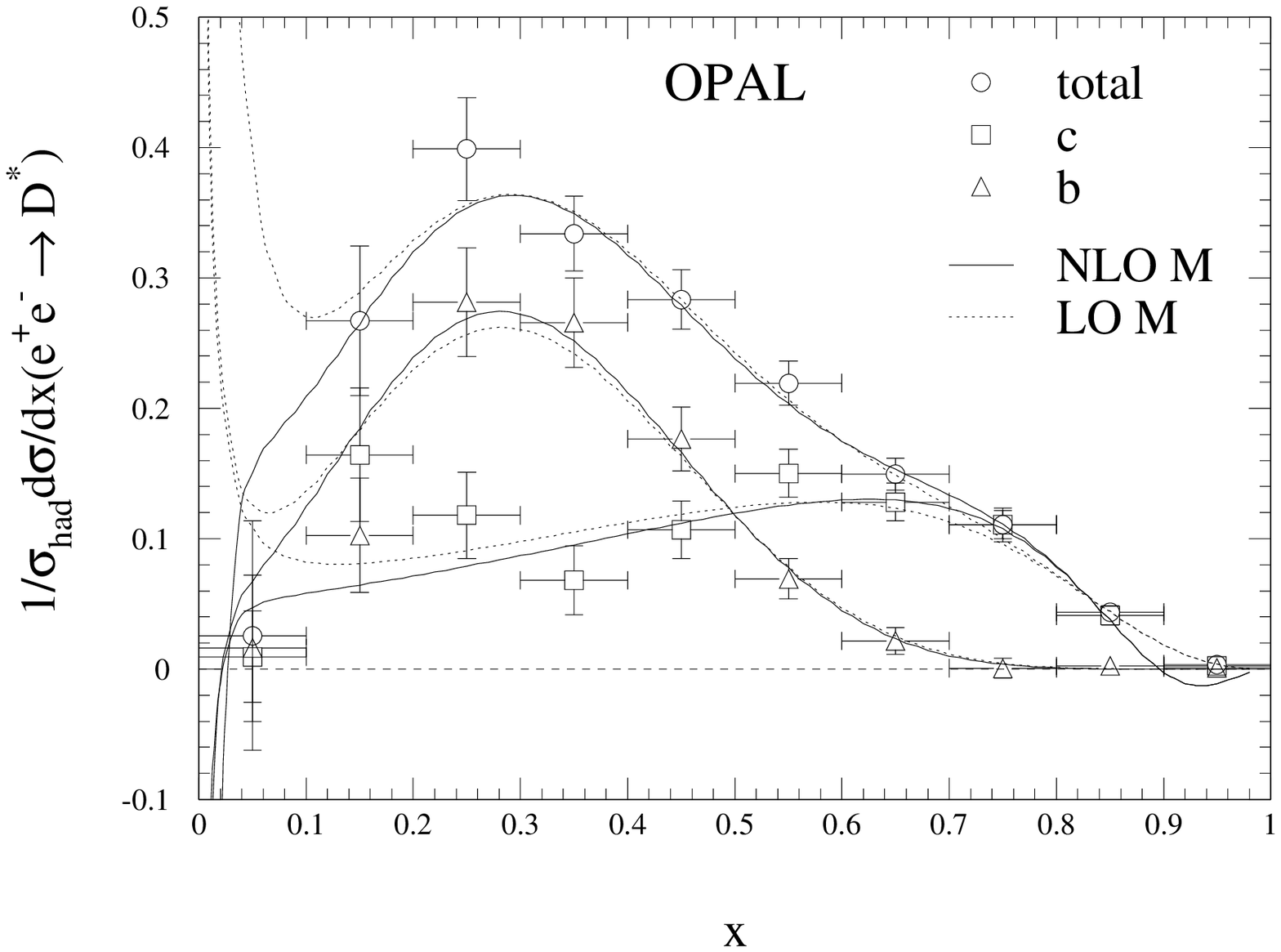,width=\textwidth}
\centerline{\Large\bf Fig.~2b}
\end{figure}

\newpage
\begin{figure}[ht]
\epsfig{figure=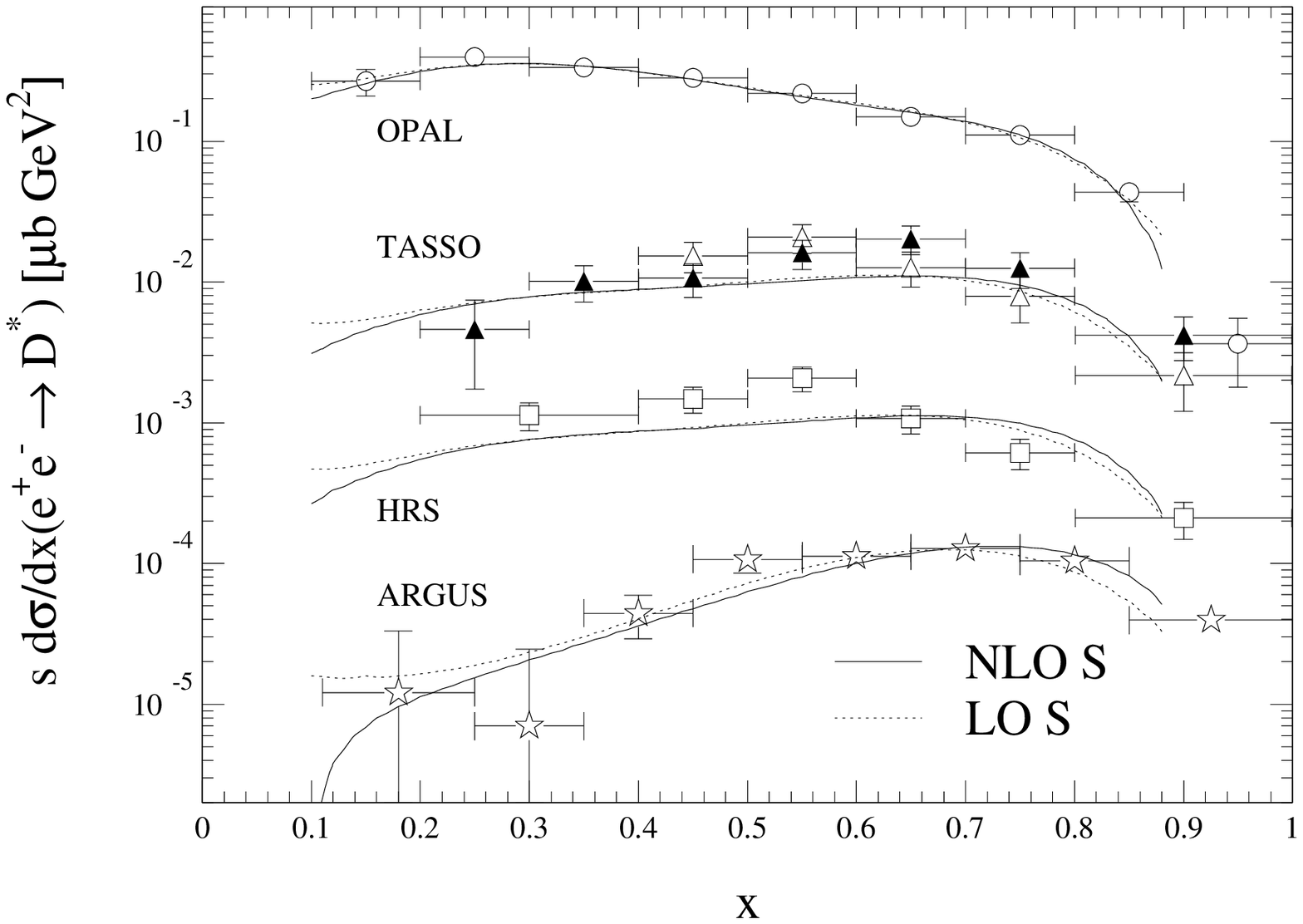,width=\textwidth}
\centerline{\Large\bf Fig.~3}
\end{figure}

\newpage
\begin{figure}[ht]
\epsfig{figure=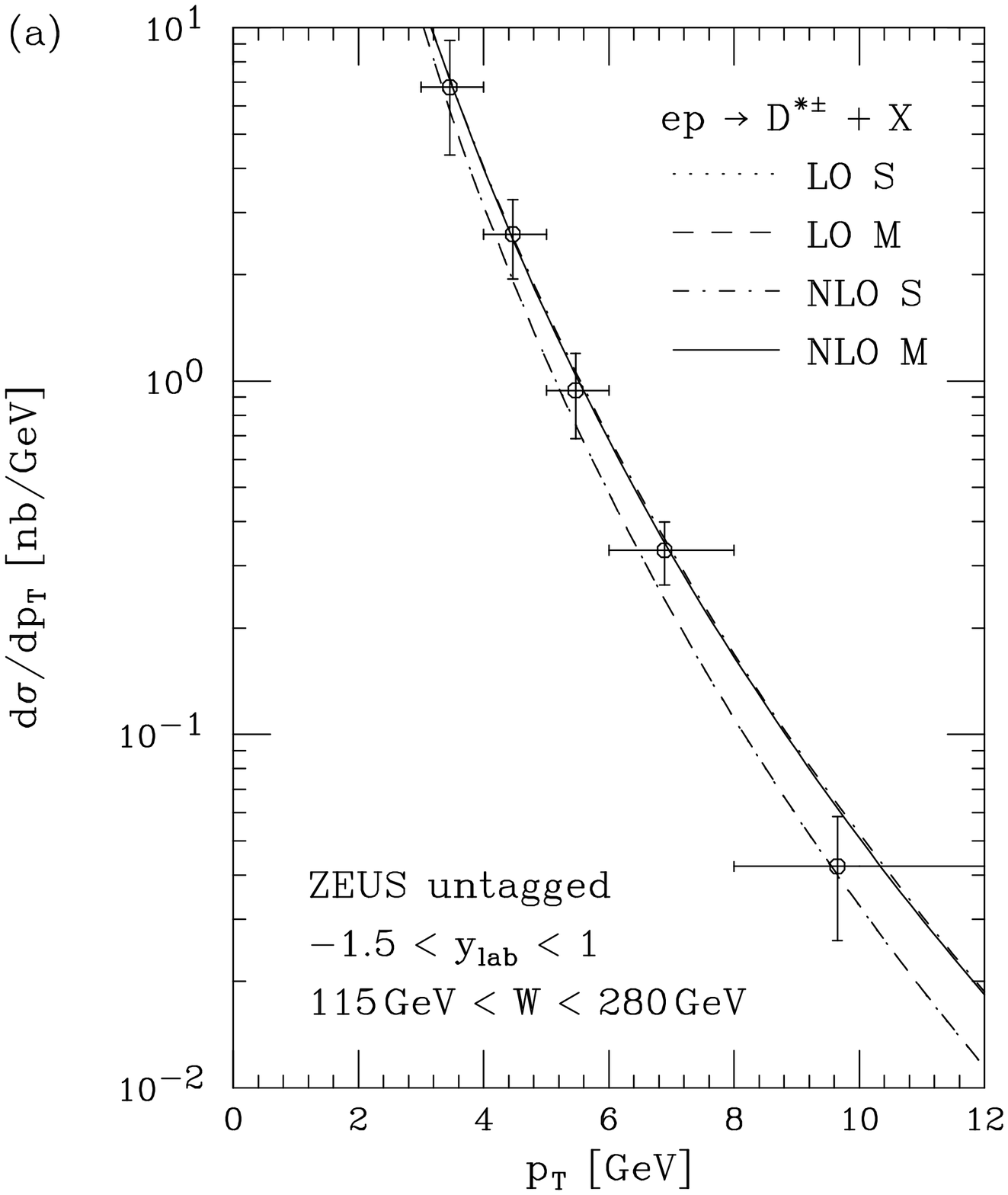,width=\textwidth}
\end{figure}

\newpage
\begin{figure}[ht]
\epsfig{figure=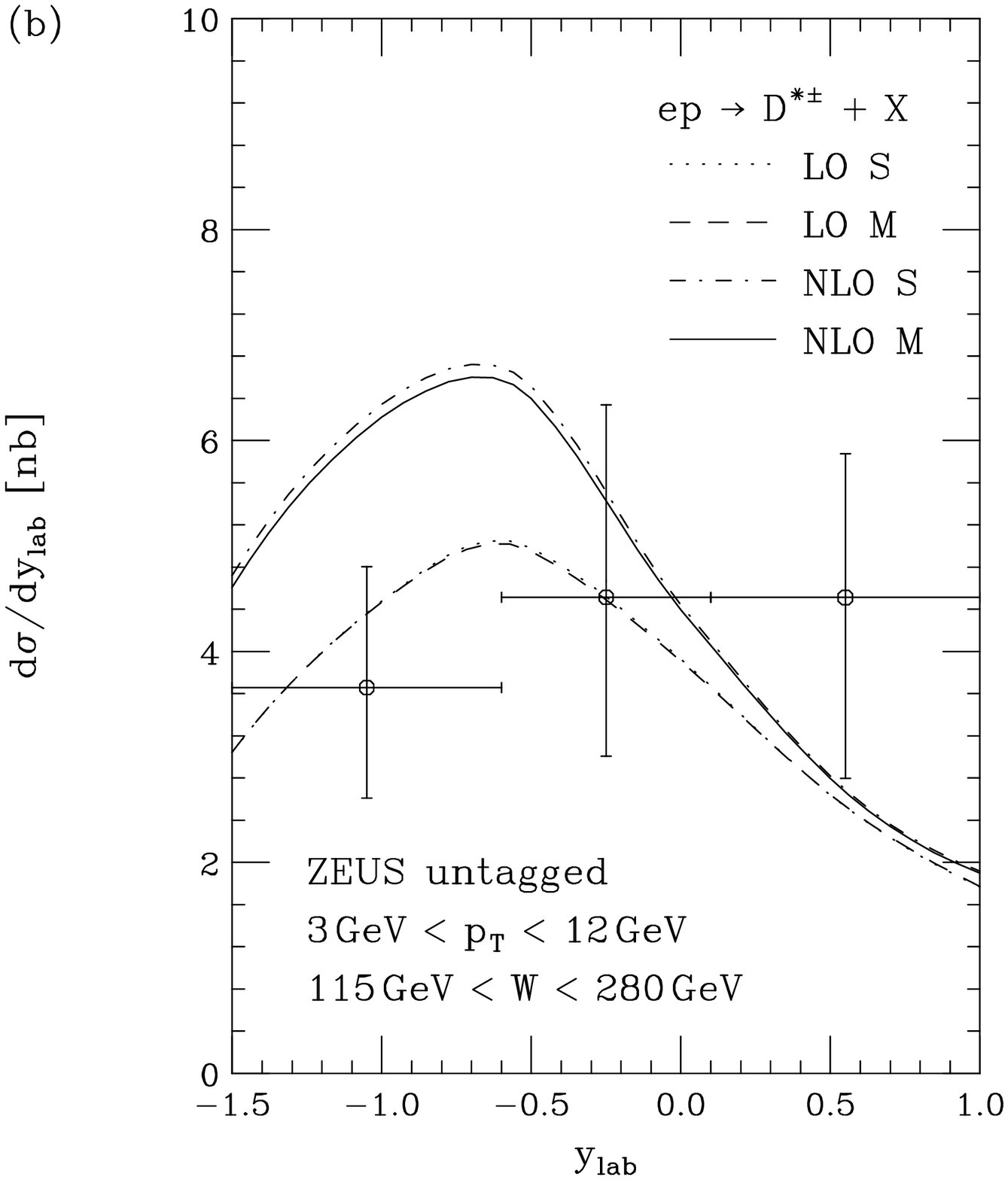,width=\textwidth}
\end{figure}

\newpage
\begin{figure}[ht]
\epsfig{figure=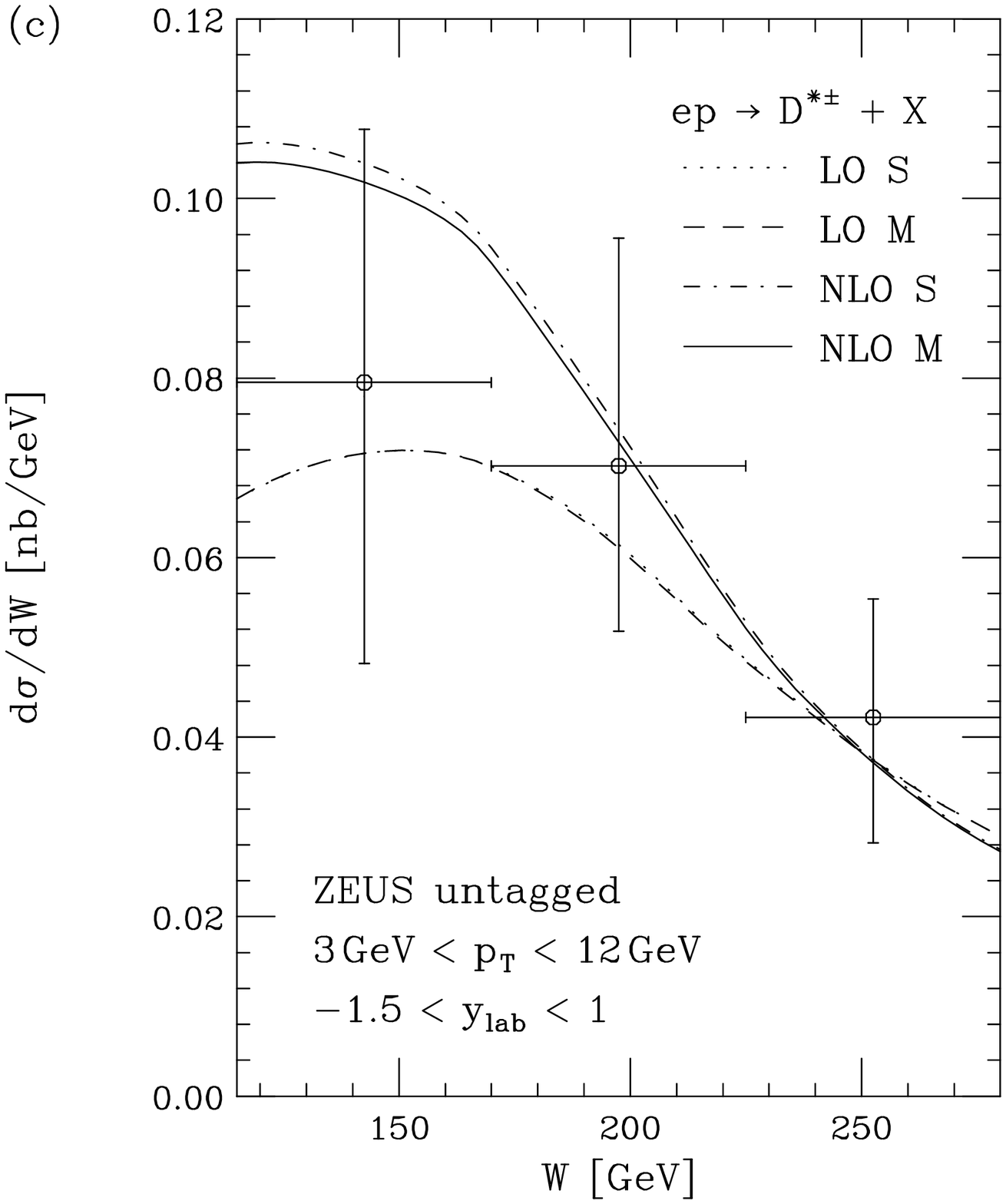,width=\textwidth}
\end{figure}

\newpage
\begin{figure}[ht]
\epsfig{figure=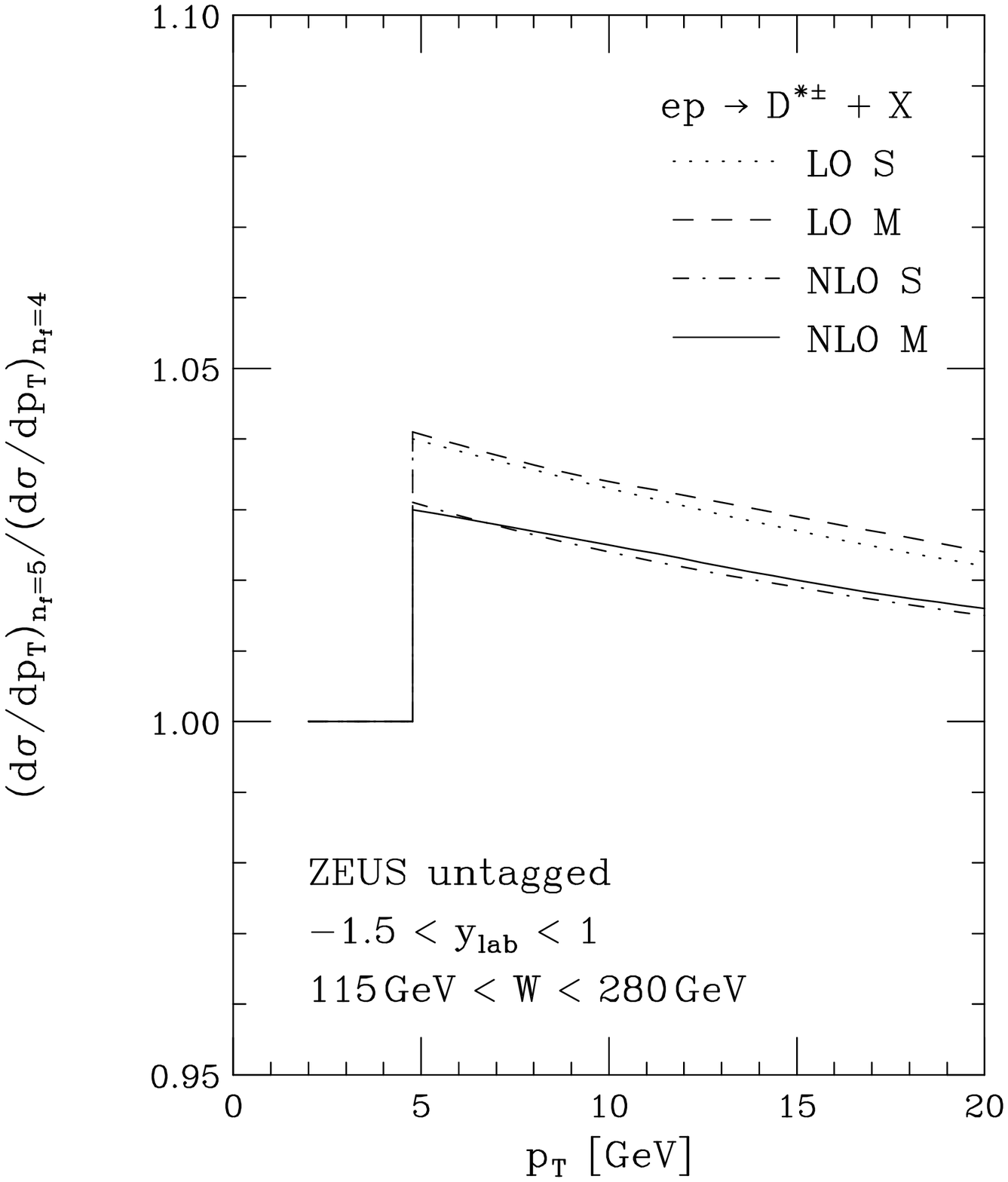,width=\textwidth}
\end{figure}

\newpage
\begin{figure}[ht]
\epsfig{figure=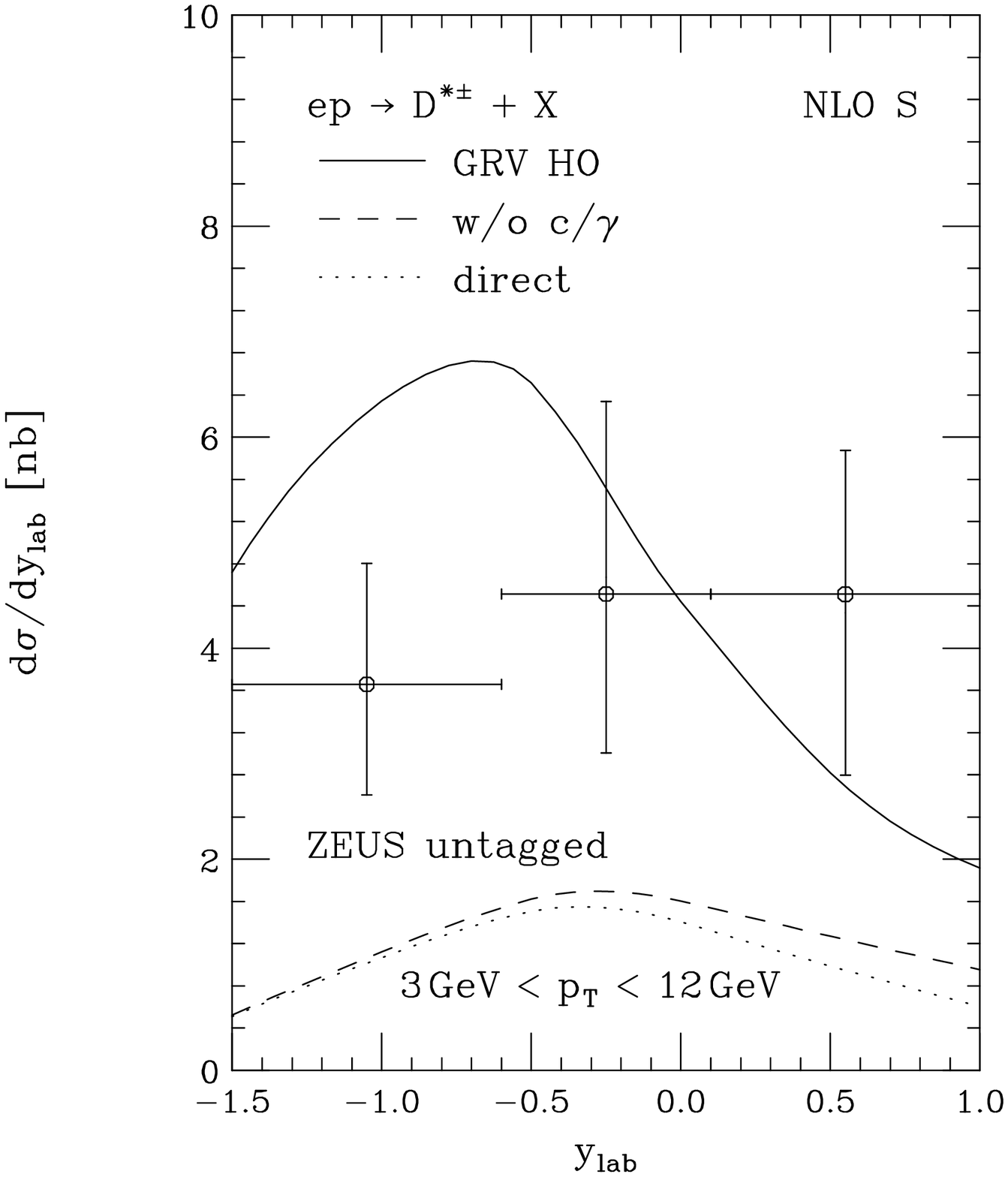,width=\textwidth}
\end{figure}

\newpage
\begin{figure}[ht]
\epsfig{figure=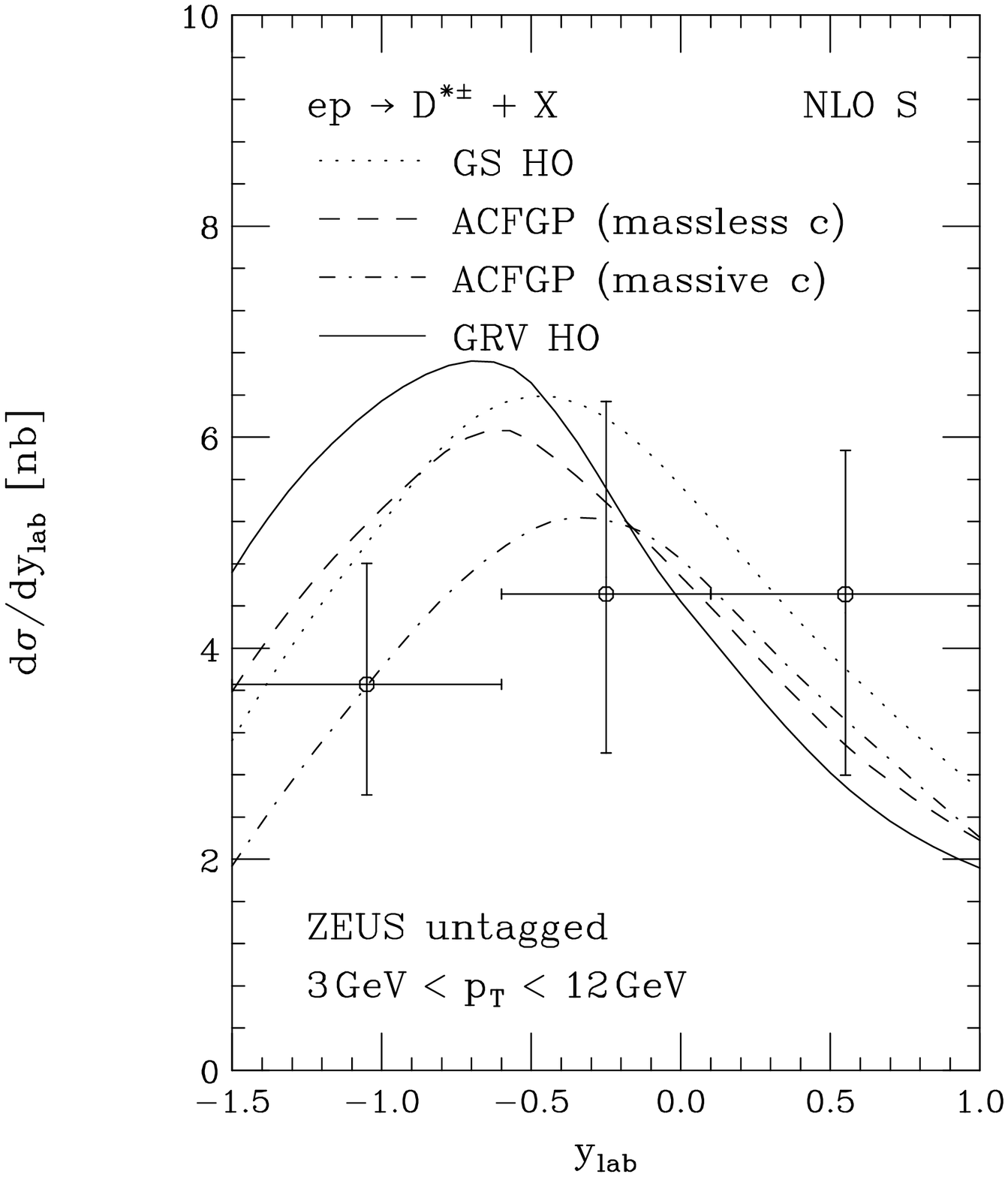,width=\textwidth}
\end{figure}


\begin{thebibliography}{99}

\bibitem{1}
H1 Collaboration, S. Aid et al.:
Nucl.\ Phys.\ B472, 32 (1996)

\bibitem{2}
ZEUS Collaboration, J. Breitweg et al.:
Phys.\ Lett.\ B401, 192 (1997);
C. Coldewey: private communication

\bibitem{3}
B.A. Kniehl, G. Kramer, M. Spira:
Report Nos.\ DESY 96--210, CERN--TH/96--274, MPI/PhT/96--103, and
hep--ph/9610267 (October 1996), Z. Phys.\ C (in press)

\bibitem{4}
P. Nason, S. Dawson, R.K. Ellis:
Nucl.\ Phys.\ B327, 49 (1989)

\bibitem{5}
B. Lampe:
Fortschr.\ Phys.\ 40, 329 (1992);
M. Cacciari, M. Greco:
Nucl.\ Phys.\ B421, 530 (1994)

\bibitem{7}
B.A. Kniehl, M. Kr\"amer, G. Kramer, M. Spira:
Phys.\ Lett.\ B356, 539 (1995)

\bibitem{8}
M. Cacciari, M. Greco:
Z. Phys.\ C69, 459 (1996)

\bibitem{9}
M. Cacciari, M. Greco, B.A. Kniehl, M. Kr\"amer, G. Kramer, M. Spira:
Nucl.\ Phys.\ B466, 173 (1996)

\bibitem{10}
TOPAZ Collaboration, R. Enomoto et al.:
Phys.\ Lett.\ B328, 535 (1994)

\bibitem{11}
M. Gl\"uck, E. Reya, A. Vogt:
Phys.\ Rev.\ D46, 1973 (1992)

\bibitem{12}
L.E. Gordon, J.K. Storrow:
Z. Phys.\ C56, 307 (1992)

\bibitem{13}
P. Aurenche, P. Chiapetta, M. Fontannaz, J.P. Guillet, E. Pilon:
Z. Phys.\ C56, 589 (1992)

\bibitem{14}
C. Peterson, D. Schlatter, I. Schmitt, P.M. Zerwas:
Phys.\ Rev.\ D27, 105 (1983)

\bibitem{15}
J. Chrin:
Z. Phys.\ C36, 163 (1987)

\bibitem{16}
ALEPH Collaboration, D. Buskulic et al.:
Z. Phys.\ C62, 1 (1994)

\bibitem{17}
OPAL Collaboration, R. Akers et al.:
Z. Phys.\ C67, 27 (1995)

\bibitem{18}
For a compilation, see G.D. Lafferty, P.I. Reeves, M.G. Whalley:
J. Phys.\ G: Nucl.\ Part.\ Phys.\ 21, A1 (1995)

\bibitem{30}
TASSO Collaboration, W. Braunschweig et al.:
Z. Phys.\ C44, 365 (1989)

\bibitem{27}
HRS Collaboration, P. Barniger et al.:
Phys.\ Lett.\ B206, 551 (1988)

\bibitem{19}
ARGUS Collaboration, H. Albrecht et al.:
Z. Phys.\ C52, 353 (1991)

\bibitem{35}
G. Altarelli, R.K. Ellis, G. Martinelli, S.-Y. Pi:
Nucl.\ Phys.\ B160, 301 (1979)

\bibitem{21}
B. Mele, P. Nason:
Nucl.\ Phys.\ B361, 626 (1991)

\bibitem{22}
G. Altarelli, G. Parisi:
Nucl.\ Phys.\ B126, 298 (1977)

\bibitem{23}
J. Binnewies, B.A. Kniehl, G. Kramer:
Phys.\ Rev.\ D52, 4947 (1995)

\bibitem{24}
G. Curci, W. Furmanski, R. Petronzio:
Nucl.\ Phys.\ B175, 27 (1980);
W. Furmanski, R. Petronzio:
Phys.\ Lett.\ 97B, 437 (1980);
P.J. Rijken, W.L. van Neerven:
Nucl.\ Phys.\ B487, 233 (1997);
M. Stratmann, W. Vogelsang:
Nucl.\ Phys.\ B496, 41 (1997)

\bibitem{25}
P. Colas: private communication

\bibitem{32}
Particle Data Group, L. Montanet et al.:
Phys.\ Rev.\ D50, 1173 (1994)

\bibitem{32a}
Particle Data Group, R.M. Barnett et al.:
Phys.\ Rev.\ D54, 1 (1996)

\bibitem{33}
H.L. Lai, J. Huston, S. Kuhlmann, F. Olness, J. Owens, D. Soper, W.K. Tung,
H. Weerts:
Phys.\ Rev.\ D55, 1280 (1997)

\bibitem{34}
J. Binnewies, M. Erdmann, B.A. Kniehl, G. Kramer:
in Future Physics at HERA: Proceedings of the Workshop 1995/96,
edited by G. Ingelman, A. De Roeck, R. Klanner, Vol.~1, p.~549

\bibitem{13a}
M. Gl\"uck, K. Grassie, E. Reya:
Phys.\ Rev.\ D30, 1447 (1984)

\bibitem{36}
M. Cacciari, M. Greco:
Phys.\ Rev.\ D55, 7134 (1997)

\end{thebibliography}
\end{document}